\documentclass[journal,transmag]{IEEEtran}

\usepackage{hyphenat}

\usepackage{titlesec}

\usepackage{ragged2e}
\justifying

\usepackage{titlesec}
\titleformat{\section}{\normalfont\large\bfseries}{\thesection}{1em}{}
\titleformat{\subsection}{\normalfont\normalsize\bfseries}{\thesubsection}{1em}{}
\titleformat{\subsubsection}{\normalfont\normalsize\bfseries}{\thesubsubsection}{1em}{}

\renewcommand{\thesection}{\arabic{section}}
\renewcommand{\thesubsection}{\thesection.\arabic{subsection}}
\renewcommand{\thesubsubsection}{\thesubsection.\arabic{subsubsection}}
\titleformat{\section}{\normalfont\large\bfseries}{\thesection}{1em}{}
\titleformat{\subsection}{\normalfont\normalsize\bfseries}{\thesubsection}{1em}{}
\titleformat{\subsubsection}{\normalfont\normalsize\bfseries}{\thesubsubsection}{1em}{}
\renewcommand{\thetable}{\arabic{table}}

\usepackage{lipsum}  

\usepackage[utf8]{inputenc}
\usepackage{comment}
\usepackage{siunitx}
\usepackage[nolist]{acronym}

\usepackage{gensymb}
\usepackage{multirow}
\usepackage{tabularray}
\usepackage{xcolor}

\definecolor{VALOSgreen}{RGB}{113, 178, 42} 
\definecolor{VALOSred}{rgb}{0.6350, 0.0780, 0.1840}
\definecolor{WO}{rgb}{0.6745, 0.9765, 0.5686}
\definecolor{WA}{rgb}{1.0000, 0.9020, 0.6000}
\definecolor{WW}{rgb}{0.9569, 0.6941, 0.5137}
\definecolor{WS}{rgb}{0.9373, 0.4588, 0.5255}
\definecolor{Gray}{gray}{0.85}

\usepackage{soul} 

\usepackage{graphicx}
\usepackage{caption}

\ifCLASSOPTIONcompsoc
\usepackage[caption=false,font=normalsize,labelfont=sf,textfont=sf]{subfig}
\else
\usepackage[caption=false,font=footnotesize]{subfig}
\fi

\ifCLASSINFOpdf

\else

\fi

\usepackage{amsmath}

\usepackage{algorithmic}

\usepackage{array}

\usepackage{hyperref}
\usepackage{soul}
\usepackage{cite}

\usepackage{todonotes}
\usepackage{float}

\begin{acronym}
	\acro{uav}[UAV]{unmanned aerial vehicle}
	\acroplural{uav}[UAVs]{unmanned aerial vehicles}
	\acro{uam}[UAM]{Urban Air Mobility}
	\acro{utm}[UTM]{Unmanned Aircraft System Traffic Management}
	\acro{vll}[VLL]{very low level}
	\acro{sdr}[SDR]{software defined radio}
	\acro{a2a}[A2A]{air-to-air}
	\acro{a2g}[A2G]{air-to-ground}
	\acro{rf}[RF]{radio frequency}
	\acro{pts}[PTS]{Portable Transmitting Station}
	\acro{rft}[RFT]{Radio Frequency Tuner}
	\acro{dru}[DRU]{Digital Receiving Unit}
	\acro{los}[LOS]{line of sight}
	\acroplural{wsn}[WSNs]{wireless sensor networks}
	\acro{fanet}[FANET]{flying ad-hoc network}
	\acroplural{fanet}[FANETs]{flying ad-hoc networks}
	\acro{cnpc}[CNPC]{control and non payload communication}
	\acro{agc}[AGC]{automatic gain control}
	\acro{gnss}[GNSS]{global navigation satellite system}
	\acro{d2d}[D2D]{Drone-to-Drone}
	\acro{kest}[KEST]{Kalman enhanced super resolution tracking}
	\acro{mpc}[MPC]{multipath component}
	\acroplural{mpc}[MPCs]{multipath components}
	\acro{pdf}[PDF]{probability density function}
	\acro{kest}[KEST]{Kalman enhanced super resolution tracking algorithm}
	\acro{gscm}[GSCM]{geometrical based stochastic channel model}
	\acro{gbas}[GBAS]{ground based augmentation system}
	\acro{cots}[COTS]{commercially-off-the-shelf}
	\acro{snr}[SNR]{Signal to Noise Ratio}
	\acro{dronecast}[DroneCAST]{Drone Communication and Suriveillance Technology}
\end{acronym}

\makeatletter
\def\endthebibliography{%
	\def\@noitemerr{\@latex@warning{Empty `thebibliography' environment}}%
	\endlist
}
\makeatother

\begin{document}
	\bstctlcite{IEEEexample:BSTcontrol}
%
\title{Towards Robust and Efficient Communications\\ for Urban Air Mobility}


\author{\IEEEauthorblockN{Dennis Becker\IEEEauthorrefmark{1},
Lukas Schalk\IEEEauthorrefmark{1},
}
\IEEEauthorblockA{\IEEEauthorrefmark{1}Institute of Communications and Navigation,
German Aerospace Center (DLR), Wessling, Germany}
}

\IEEEtitleabstractindextext{%
\begin{abstract}

\section*{\textbf{Abstract}}
For the realization of the future urban air mobility, reliable information exchange based on robust and efficient communication between all airspace participants will be one of the key factors to ensure safe operations. Especially in dense urban scenarios, the direct and fast information exchange between drones based on Drone-to-Drone communications is a promising technology for enabling reliable collision avoidance systems. However, to mitigate collisions and to increase overall reliability, unmanned aircraft still lack a redundant, higher-level safety net to coordinate and monitor traffic, as is common in today's civil aviation. In addition, direct and fast information exchange based on ad hoc communication is needed to cope with the very short reaction times required to avoid collisions and to cope with the the high traffic densities. Therefore, we are developing a \ac{d2d} communication and surveillance system, called DroneCAST, which is specifically tailored to the requirements of a future urban airspace and will be part of a multi-link approach. In this work we discuss challenges and expected safety-critical applications that will have to rely on communications for \ac{uam} and present our communication concept and necessary steps towards DroneCAST. As a first step towards an implementation, we equipped two drones with hardware prototypes of the experimental communication system and performed several flights around the model city to evaluate the performance of the hardware and to demonstrate different applications that will rely on robust and efficient communications.
\end{abstract}

\begin{IEEEkeywords}
unmanned aviation, urban air mobility, drone-to-drone communications, collision avoidance, measurements, flight demonstration
\end{IEEEkeywords}}

\maketitle

\IEEEdisplaynontitleabstractindextext

\IEEEpeerreviewmaketitle


\section*{NOMENCLATURE}

\resizebox{\columnwidth}{!}{
\begin{tabular}{@{}ll}
AGC & Automatic Gain Control\\
CNPC & Control and Non Payload Communication\\
COTS & Commercially off the Shelf\\
DAA & Detect and Avoid\\
DroneCAST & Drone Communication and Surveillance Technology\\
D2D & Drone-to-Drone\\
GBAS & Ground Based Augmentation System\\
GPSDO & GPS Disciplined Oscillator\\
LOS & Line of Sight\\
SDR & Software Defined Radio\\
SNR & Signal to Noise Ratio\\
UAM & Urban Air Mobility\\
UTM & Unmanned Aircraft System Traffic Management
\end{tabular}
}


\section{Introduction}
In the near future, the urban airspace will be shared by piloted as well as unpiloted and autonomous aircraft, so-called drones. Current airspace management concepts, such as SESAR U-Space~\cite{2017sesarjointundertakingUspaceBlueprint} and NASA UTM~\cite{2020faa/nasaUTMConceptOperations}, rely on a reliable exchange of information between all participants for a safe integration of the new participants in urban airspace. In particular, unpiloted aircraft such as drones depend on this data exchange. Although robust communication is a central aspect of all concepts, there is currently no communication system that has been adapted to the specific challenges of this environment. In addition, due to the high density of drones, the management of urban airspace, called \ac{utm}, will be fundamentally different from the way it is currently handled in civil aviation. Continuous remote control of all the drones by a remote pilot in communication with UTM will not be possible due to the high traffic density and short reaction times needed to avoid collisions. Instead, UTM will heavily rely on pre-planned and conflict-free trajectories as well as continuous monitoring. Drones will fly these trajectories in an automated or autonomous manner. The implementation of this UTM concept will rely, at least in part, on existing communications infrastructure, such as mobile communication in order to connect the drones to the UTM \cite{2017geisterDLRUSpaceBlueprint} . Under ideal conditions, this approach may seem sufficient. However, upon closer inspection, weaknesses quickly become apparent, such as a lack of redundancy or a lack of an overarching safety net, as is common in civil aviation and shipping, or as is envisioned for future autonomous driving \cite{2009rtcaDO260BMinimumOperational,2011rtcaDO282BCorrigendumMinimum,2014ituTechnicalCharacteristicsAutomatic,201080211p2010IEEE} . \\
 But the urban environment is very challenging from a physical layer point of view, with rich multipath signal propagation as well as shadowing and diffraction events when flying close to surrounding objects such as tall buildings.
Therefore, we are developing an ad hoc communication concept that is adapted to the specific challenges of the urban environment and takes into account the requirements of the potential applications. The ad hoc communication concept refers to the technical communication on the air interface between different nodes and is designed as a redundant data link in addition to other communication options in the context of a multi-link approach.


%


\section{Challenges and Applications for Communication Systems for Urban Air Mobility}
\begin{figure}[H]
	\centering
	\includegraphics[width=0.98\columnwidth]{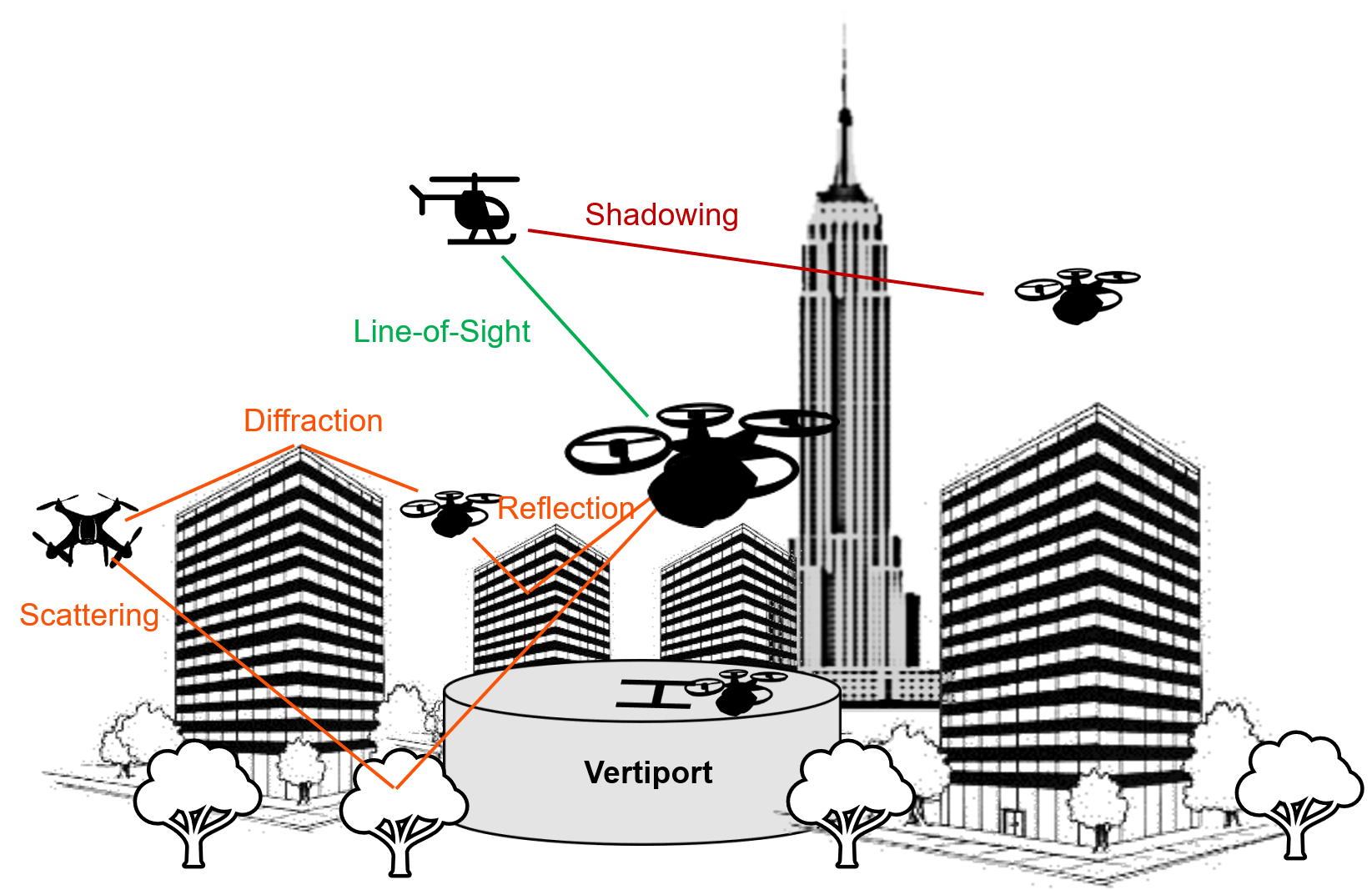}
	\caption{Major signal propagation effects to consider in the urban D2D communications channel.}
	\label{fig:urban_channel}
\end{figure}

\begin{figure*}[ht!]
	\centering
	\includegraphics[width=\textwidth,height=8cm]{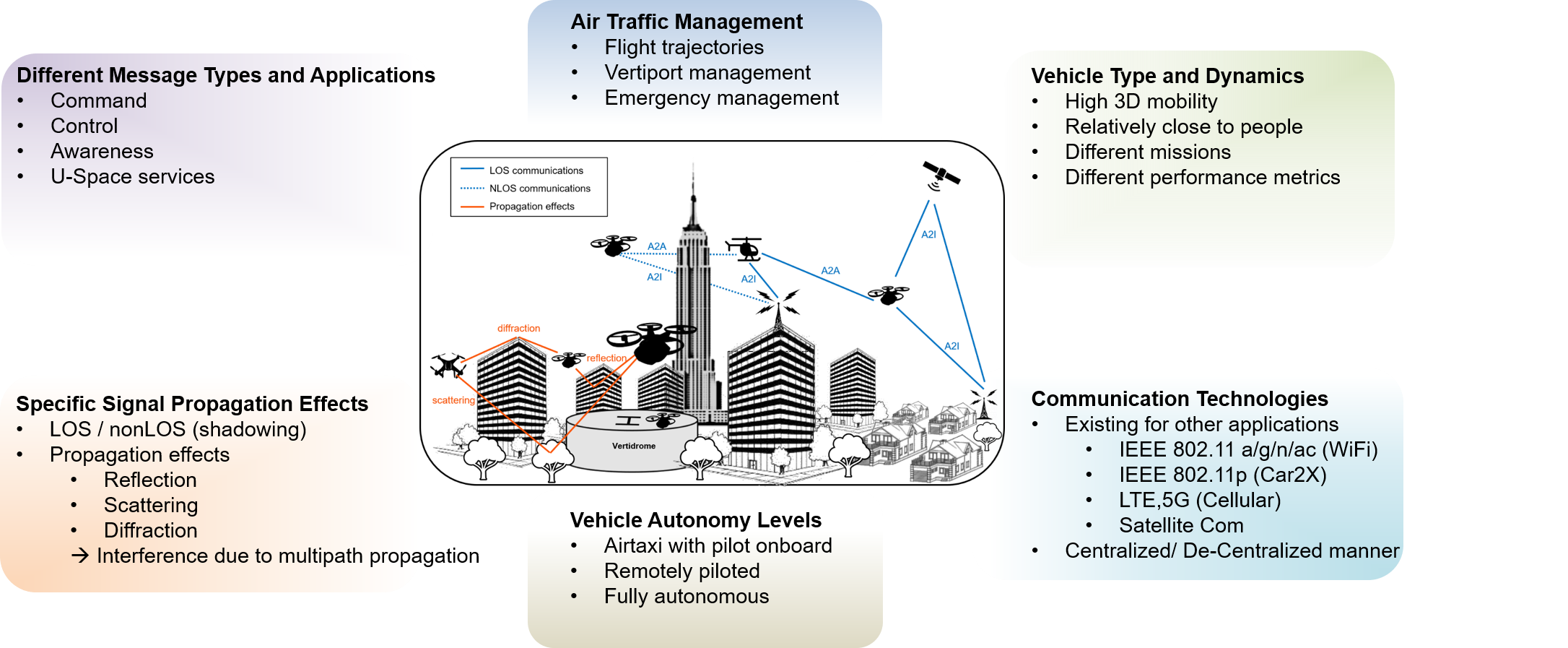}
	\caption{Overview of categories that have direct or indirect mutual relationships with communication for Urban Air Mobility.}
	\label{fig:topics}
\end{figure*}

Communication systems for use in urban airspace face unique challenges that must be considered when selecting an appropriate system. The expected high density of drones must be considered along with high mobility in three-dimensional space and rapidly changing topologies. Communication resources are limited and must be shared by all participants, whether airborne or ground based. The efficient use of resources and scalability is critical. In urban environments, the transmitted electromagnetic signals are reflected, scattered and diffracted by many surrounding objects such as buildings, vegetation and cars like illustrated in fig.~\ref{fig:urban_channel} The multipath propagation of the signal can cause unfavorable overlap at the receiver and must be taken into account during reception to allow reconstruction of the transmitted signal. In addition, such interference can also be expected between different signals of the participants, especially in the air, where there is a high visibility between the vehicles and possible communication infrastructures in a dense space \cite{2016vanderberghLTESkyTrading}. In addition, direct signal propagation can be expected to be shadowed by larger objects such as buildings at lower altitudes, so that only reflected and diffracted components can be received. Influences from the aircraft itself, such as shadowing from their frame, electrical and mechanical sources of interference, must also be considered.
The layout of the aircraft's capability may also lead to possible limitations in terms of size, weight and power consumption, known as SWaP constraints. This must then be taken into account in the choice of the transceiver performance. In addition, the requirements of various applications and future regulations in the field of urban air mobility are not yet well known. Figure~\ref{fig:topics} provides an overview of several other categories that interact with communications in the urban airspace and may need to be considered.
In addition to the specific signal propagation effects already mentioned, any application that requires information exchange with the airborne vehicles may also place direct requirements on the communication system. For example, a certain amount of data must be transmitted within a certain time, the data link must be highly available, or a minimum number of subscribers, i.e., scalability, must be ensured. There may also be indirect requirements or influences on the communication system. For example, a command or important information for an aircraft to avoid an obstacle may need to be provided in a more timely manner if the aircraft is traveling faster, if the aircraft is very sluggish in the evasion obstacles, or has a limited mobility. Also, if position accuracy is degraded, increased separation and other separation rules may need to be applied, requiring more timely or frequent information exchanges. An on-board autonomy system may need to send information back to \ac{utm}, depending on the level of autonomy, or require certain clearances that may not be automated. Required security measures also mean increased data exchange and data volume. More broadly, various environmental factors place demands on the communications system. Weather, for example, can degrade signal propagation conditions or affect flight performance, such as in the case of strong wind gusts. In addition, events such as bird strikes may require active detection and transmission of critical information to the aircraft. 

However, for communication in urban environments, collision avoidance in densely populated airspace will be a key application, as reliable and decentralized exchange of position data and trajectories between individual drones will be required. In this context, there is a high demand for the lowest possible transmission latency to enable the shortest possible reaction times.


\section{Multi-link Approach for Robust Communication} \label{sec:multilink}
To realize the upcoming \ac{uam}, a wide variety of applications will be used, each with different requirements on communication. It is therefore very difficult for a single communication system to cover the wide range of requirements. Therefore, we pursue a multi-link approach, i.e. a combination of different data links, as it is also aimed at in other concepts \cite{2020faa/nasaUTMConceptOperations,2019kovacsDroC2om763601D4,2022fraunhoferheinrich-hertz-institutSUCOMSuperiorUTM,2022dauerAutomatedLowAltitudeAir}. A multi-link approach combining different data link technologies has many advantages over a single data link. The increase in redundancy and the increase in the performance of the overall system are the key aspects here. In addition, the initial effort required for the step-by-step implementation of applications such as U-Space Services is reduced. Existing communication systems can be used first, even if they have not been adapted for the application, and then future adapted data links can be added or the existing systems can be adapted according to the requirements. We distinguish different systems in the categories of infrastructure communication and adhoc communication. 

For most use cases, the already existing communication infrastructure in the urban area will be sufficient, since for a large part the exchange of information is not security-critical and the required amounts of data can be transmitted over it with a certain delay. Since collision avoidance is a particularly safety-critical application in urban areas, we consider an specifically tailored and redundant safety network based on an ad hoc communication system to be the most important element for this application. In this case, important information for collision avoidance should be exchanged primarily via a direct and adapted data link between vehicles. 

\begin{figure}[H]
	\centering
	\includegraphics[width=0.98\columnwidth]{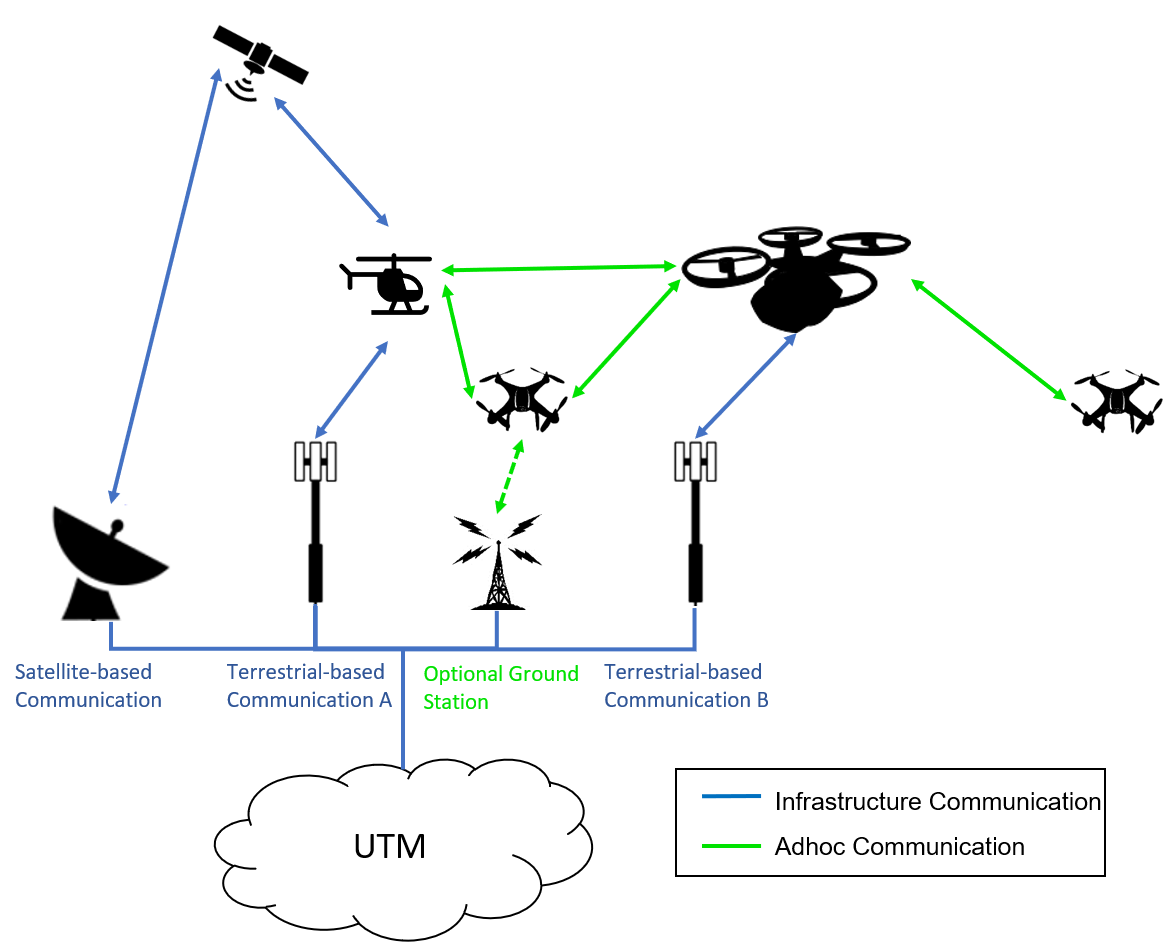}
	\caption{Multi-link approach as communication concept in DLR project HorizonUAM.}
	\label{fig:comconcept_huam}
\end{figure}

We consider the construction of a combined communication and monitoring system, which shall meet the following characteristics.\\
\begin{enumerate}
	\item \textbf{Cooperative Collision Avoidance}\\
	Cooperative collision avoidance based on ad hoc communication between drones will be implemented that creates an additional, decentralized safety net without having to rely on communication infrastructure.\\
	
	\item \textbf{Redundant Monitoring and Tracking of Aircraft}\\
	Not only can ad hoc communications be used to establish a direct link between drones, but redundant monitoring of drone movements can be established using appropriate ground stations to support the \ac{utm}. This can be done using the position messages of all drones in range, which are already broadcast for collision avoidance.\\
	
	\item \textbf{Backup Datalink}\\
	For most applications, non-critical information can be sent over existing communication infrastructure. Nevertheless, it may be useful to have a redundant bi-directional data link available for this as well to increase reliability. Furthermore, it is not yet possible to estimate which other possible critical applications will require a reliable data link or low latencies in addition to collision avoidance. Therefore, it makes sense to build a basic backup data link that goes beyond a conventional pure beaconing system for collision avoidance. Here it would be possible to establish only direct links between the airborne participants and possible ground stations or to allow links over multiple "hops". A "multihop" communication across multiple participants would significantly increase the capability of a backup data link and create a new communication infrastructure, but implies additional effort in implementation and overhead in the communication itself. For example, routing algorithms would need to be implemented to route messages to the recipient via the correct path. The challenge here lies primarily in the rapidly changing network topology due to the high mobility of participants and changing signal propagation conditions. Thus, an initially preferred connection path between two nodes as part of a route can quickly become unfavorable or even fail completely if the communication characteristics deteriorate or the connection is quickly disrupted by shadowing.
\end{enumerate}
Figure~\ref{fig:comconcept_huam} illustrates such an multi-link approach considering mobile communication and sat-based communication as available infrastructure in urban environments. 

\section{Adhoc Communication as Solution for Collision Avoidance in Urban Airspace}

In road traffic, drivers avoid collisions with other vehicles by using their eyes to monitor their surroundings and braking or swerving as soon as they detect that another vehicle is on a collision course. In today's vehicles, optional assistance systems help the driver detect potential collisions. For example, adaptive cruise control systems use on-board sensors such as RADAR, LIDAR, or cameras to adjust the vehicle's speed to maintain a safe distance from vehicles ahead. 

\begin{figure}[H]
	\centering
	\includegraphics[width=0.99\columnwidth]{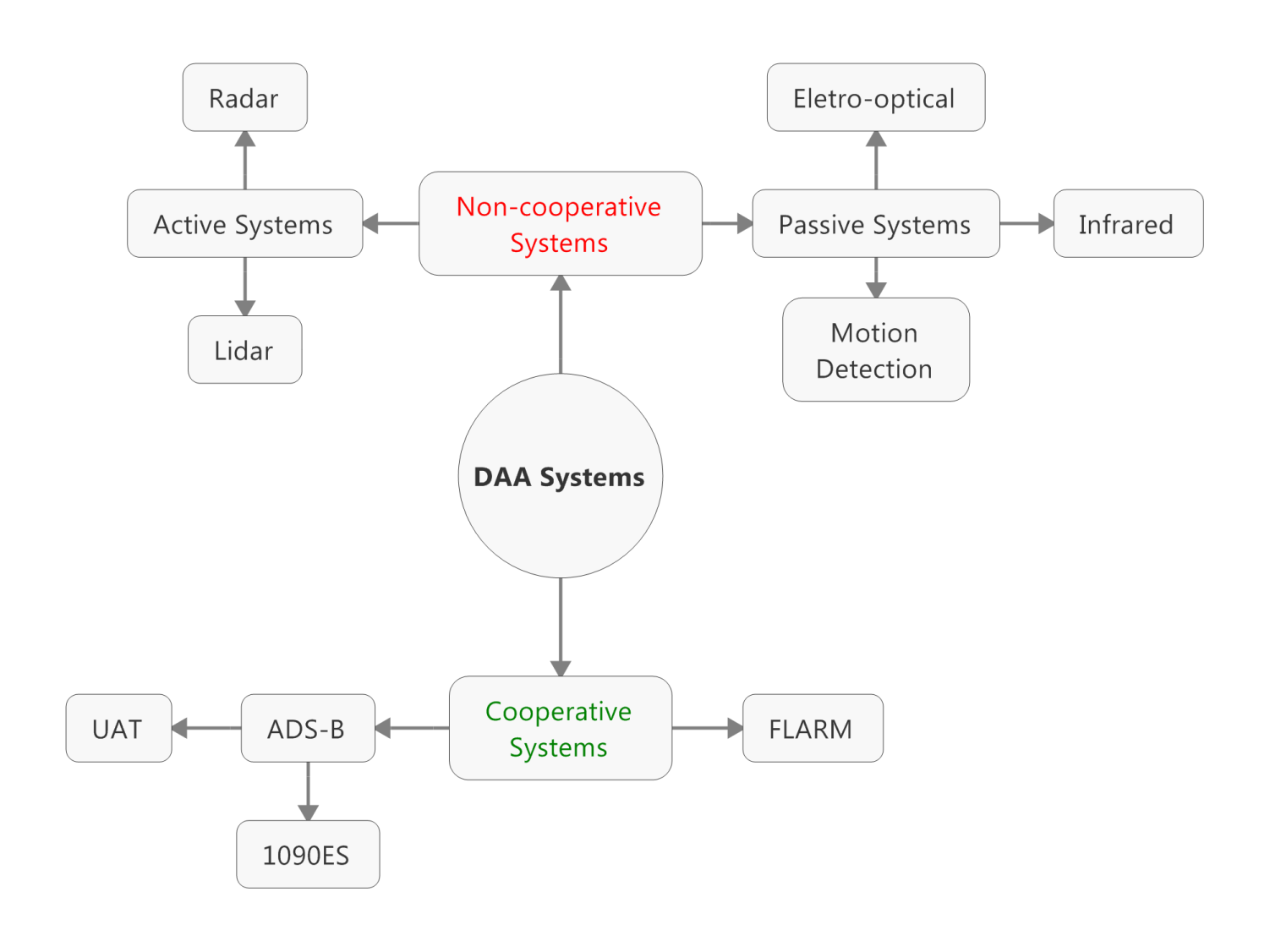}
	\caption{Overview and classification of Detect-and-Avoid systems.}
	\label{fig:daa_systems}
\end{figure}

Beyond that, a variety of other sensors can be used to detect collision courses in road traffic, air traffic, rail traffic, or maritime traffic. Basically, sensors can be divided into two types: Cooperative obstacle detection sensors and non-cooperative obstacle detection sensors. Figure~\ref{fig:daa_systems} provides an overview of the different sensor types for detect-and-avoid (DAA) systems.
Cooperative obstacles actively attract attention, for example by emitting a signal. Non-cooperative obstacles do not call attention to themselves. Sensors that detect non-cooperative systems can be further divided into active and passive systems. Active systems emit a signal and detect the reflection of the obstacle. An example of an active system is a RADAR. Passive systems detect obstacles by detecting unintentionally emitted signals, such as thermal radiation. Cooperative systems are widely used in all traffic domains to create situational awareness among vehicles. Therefore, every vehicle is required to periodically transmit its own position and intent to nearby vehicles via an ad-hoc communications system. Popular systems are 1090 Extended Squitter for air traffic, IEEE 802.11p for road traffic, RCAS for rail traffic and AIS for maritime traffic. 

The main advantage of cooperative systems is the fact that cooperative systems typically provide very accurate information about position, direction, and speed, as well as additional information that non-cooperative systems cannot easily provide, such as the type of target and its state. 
Moreover, ad-hoc communication is a way of transmitting data without relying on any fixed infrastructure. It has several advantages for collision avoidance, especially in critical situations where every millisecond counts. Ad-hoc communication can achieve lower latency than infrastructure communication, which means that the data can be delivered faster and more reliably. Ad-hoc communication can also serve as a backup in case the infrastructure communication fails or is not available. Compared to RADAR and LIDAR, ad-hoc communication can cover higher ranges and works even in case of signal shadowing, which is when the signal is blocked or weakened by objects or weather conditions. Furthermore, ad-hoc communication can enable extended information exchange between the communicating vehicles, such as evasion instructions and information about trajectories. This can help to coordinate the actions and avoid conflicts. Finally, ad-hoc communication can have lower power consumption than RADAR, which means that it can save energy and reduce costs.

Cooperative detect and avoid via ad-hoc communication systems has also some disadvantages. First, the information that is transmitted must be trusted, since there is no guarantee that it is accurate or authentic. Therefore, security measures are needed to ensure the reliability and integrity of the communication. Second, ad-hoc communication requires an additional communication module, which adds costs, weight and power consumption to the vehicles or participants. This may affect their performance and efficiency. However, safety should not be compromised for the sake of saving resources. Third, ad-hoc communication can only detect uncooperative participants, who do not share their information or intentions with others. This means that there may be some hidden or unexpected threats that are not accounted for by the communication.


\section{Development of a Drone-to-Drone Channel Model for Urban Environments based on Measurements}
In 2019, German Aerospace Center (DLR) conducted a wideband channel sounding measurement campaign with two small hexacopters to measure drone-to-drone (D2D) propagation characteristics in an urban environment. The campaign took place at the DLR site in Oberpfaffenhofen, Germany, in three different environments with different flight trajectories including critical scenarios, where two communicating drones are not always in LOS to each other and are on a collision course. A channel sounding signal was transmitted in the C-band at 5.2~GHz with a bandwidth of 100~MHz and a transmit power of 30~dBm using omnidirectional and vertically polarized radiating antennas mounted underneath the drones. The measurement setup, hardware equipment and flight scenarios are described in detail in \cite{2019beckerEnablingAirtoAirWideband,2020beckerWidebandChannelMeasurements}.\\

\begin{figure}[H]
	\centering
	\includegraphics[width=0.98\columnwidth]{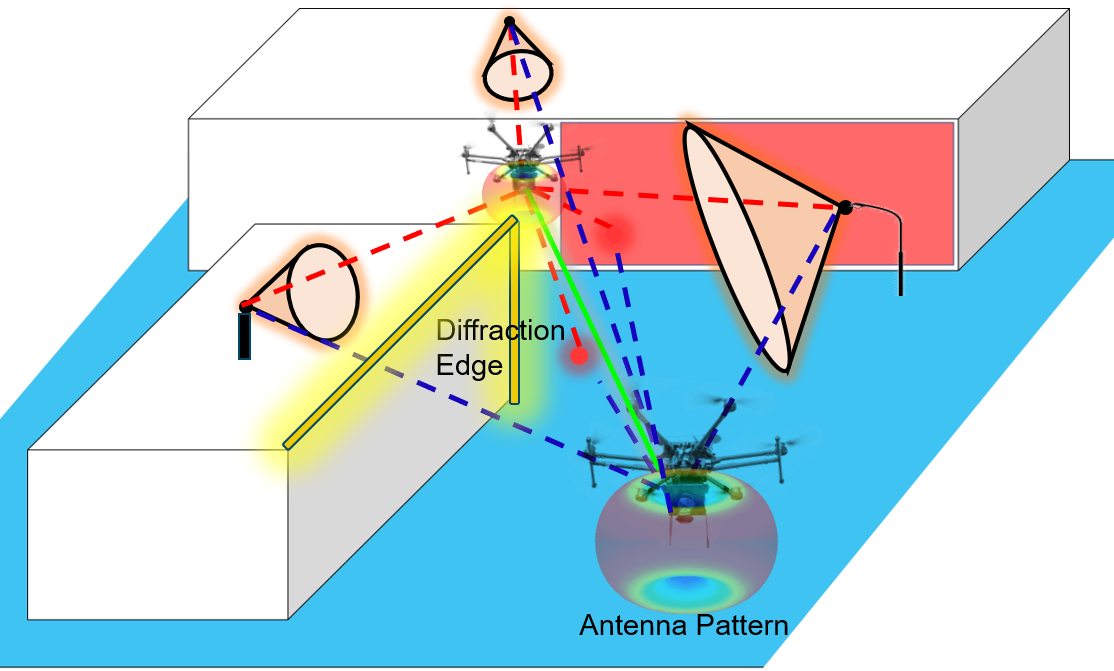}
	\caption{Overview of elements for the D2D channel model.}
	\label{fig:channel-model_architecture}
\end{figure}

\begin{figure}[H]
	\centering
	\includegraphics[width=0.98\columnwidth]{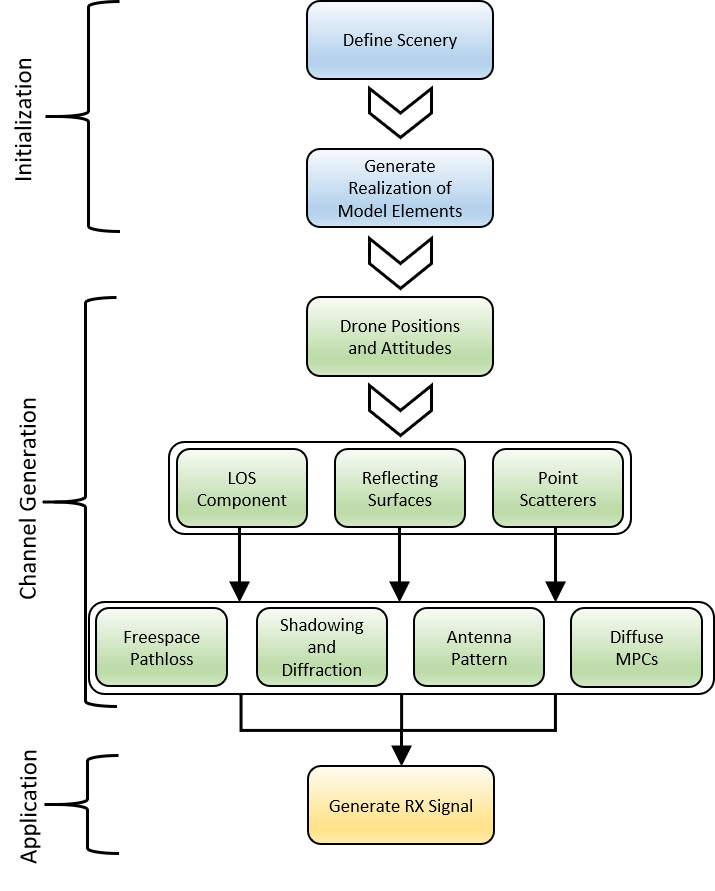}
	\caption{Overview of the simulation chain for the D2D channel model.}
	\label{fig:channel-model_sim-chain}
\end{figure}

Based on these measurements, we proposed a wideband channel model for \ac{d2d} scenarios in urban environments in \cite{2023beckerModelingDronetoDroneCommunications} to help evaluating and validating different communication concepts and datalink candidates via simulations without having the need to perform complex and time costly measurement campaigns. By considering the underlying signal propagation effects in urban environments the robustness of datalink candidates can be improved. The model follows a geometrical-statistical channel modeling (GSCM) approach and incorporates coarse-grained knowledge about realistic locations and shapes for buildings to model the propagation effects closely to their physical cause in real-world. It is antenna independent and considers the identified dominant signal propagation effects from our measurements, but can easily incorporate further statistics. A more detailed discussion on the propagation characteristics and preliminary steps are presented in \cite{2022beckerDronetoDronePropagationCharacteristicsa} and \cite{2022beckerMeasurementBasedIdentification}.
Figure~\ref{fig:channel-model_architecture} gives an overview of the model elements and fig.~\ref{fig:channel-model_sim-chain} shows the steps in the simulation chain. 
First, coarse-grained abstract building shapes are placed according to the scenario under investigation. The locations and shapes of buildings in the surrounding environment very much influence the propagation characteristics of the urban D2D channel and therefore this initial placement helps to achieve realistic distribution of all other model elements. For this, statistical descriptions for urban environments like in the ITU-R Rec. P.1410 model\cite{2012ituRECOMMENDATIONITUR14105} for example or direct 3D geometries from land surveying offices or similar can be used.
After the definition of the building shapes, the flight trajectories are defined and the properties of two different elements are drawn from statistical distributions. There are point scatterers with certain opening angles and scattering losses placed at different positions on the surfaces as well as reflection surfaces with certain dimensions and reflection losses. After this initialization phase the scenario is fully defined and the communication channel properties are generated in a snapshot based manner given the targeted time resolution. Then different propagation effects are calculated for the simplified resulting signal paths for all elements. Finally, the super-imposed signal is calculated at the receiver.



\section{Development of DroneCAST}
\begin{figure}[H]
	\centering
	\subfloat[Collision Avoidance]{\includegraphics[width=0.91\columnwidth]{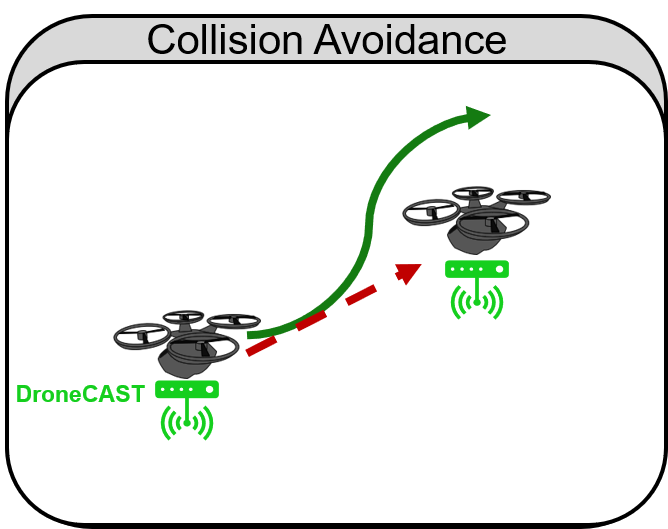}%
		\label{fig:app_ca}}
	\hfil
	\subfloat[GBAS Transmitter]{\includegraphics[width=0.91\columnwidth]{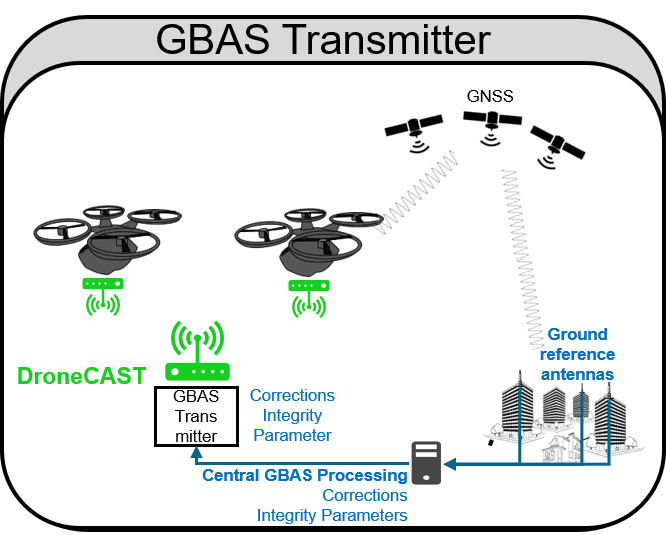}%
		\label{fig:app_gbas}}
	\hfil
	\subfloat[Vertiport Communication]{\includegraphics[width=0.92\columnwidth]{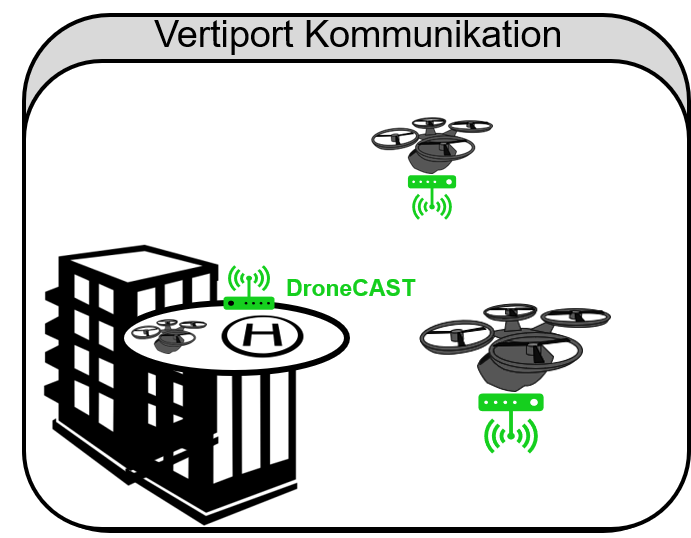}%
		\label{fig:app_vertiport}}
	\caption{Expected safety-critical applications for urban air mobility relying on communications over radio interface.}
	\label{fig:applications}
\end{figure}

As an essential part of the described multi-link approach in sec.\ref{sec:multilink}, we aim to develop a data link tailored to the special requirements and challenges for the safe operation of UAVs in urban areas. As a redundant data link based on the direct exchange between the vehicles, this supplements the communication alongside the available communication infrastructure such as through mobile or sat communication. The concept for this ad hoc communication first takes into account the requirements imposed by the main application, collision avoidance, since this is the most safety-critical application and the exchange of information is given priority. But we expect following three different applications within \ac{uam} that will have rely on communication to airspace participants over the radio and are safety-critical to ensure a safe operation. First, cooperative collision avoidance based on direct communications will be needed resolve potential conflicts in the last course of action like shown in fig.~\ref{fig:app_ca}. 
In order to ensure the high requirements for navigation such as high position accuracy and high availability, the navigation concept will also rely on redundant design of onboard sensors such as GNSS, IMU or cameras and the fusion of these sensor data. In addition to his, broadcasting GNSS correction data from \ac{gbas} groundstations to airborne vehicles may support their navigation. For this, information must be transmitted from the ground stations to the drones like shown in fig.~\ref{fig:app_gbas}. As third possible application, broadcasting important infos from vertiports to all vehicles in close distance such as in emergency cases for example might be needed like illustrated in fig~\ref{fig:app_vertiport}.

\begin{figure}[H]
	\centering
	\includegraphics[width=0.23\columnwidth]{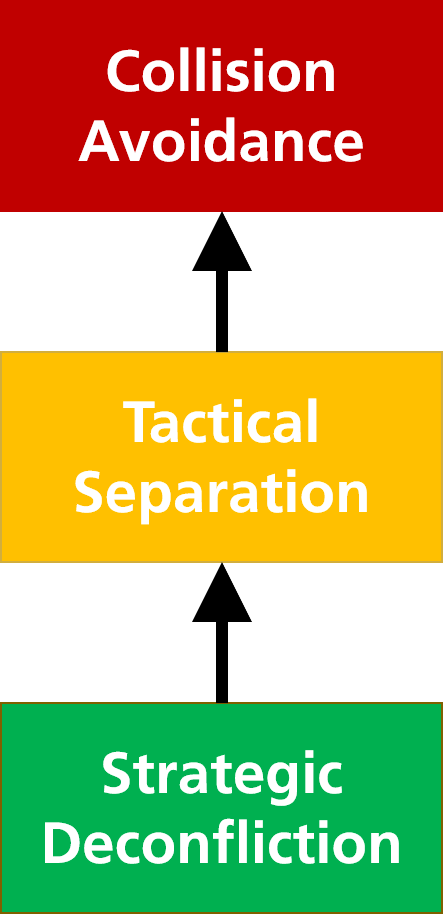}
	\caption{Three general levels of deconfliction.}
	\label{fig:deconfliction_levels}
\end{figure}

For this we proposed \ac{dronecast} \cite{2022schalkDroneCASTAnalysisRequirements} in order to establish an additional, decentralized and robust safety layer for the UTM concept as third level of the general levels like shown in fig.~\ref{fig:deconfliction_levels}. We discussed first design decisions as well as analyzed requirements for \ac{dronecast}. It is supposed to work reliably up to a drone density of 100 drones per square kilometer while using not more than 5~MHz of frequency bandwidth in the C-band at between 5030~MHz to 5091~MHz, which is already foreseen for drone communications. Major identified challenges are the severe multipath propagation environment and sudden shadowing events from a physical layer point-of-view and the high expected drone densities as well as limited communication resources from a medium access control point-of-view.

\section{Experimental Platform and Flightdemonstrations towards DroneCAST}
As a first step towards an implementation of DroneCAST, we equipped two drones with hardware prototypes of the experimental communication system and performed several flights around the model city to evaluate the performance of the hardware in comparison to \ac{cots} hardware and to demonstrate different applications that will rely on robust and efficient communications. The flight tests are to show if the hardware is suitable for a later implementation and can be flown by our drones.

\subsection{Experimental Radio}
The basis hardware for our experimental radio is a Software Defined Radio (SDR) together with a software implementation of the IEEE~802.11p WiFi standard for vehicular communication \cite{201080211p2010IEEE} on an small companion computer. The software implementation, which consists of different building blocks for a transmission system, runs in GNU radio, a signal processing framework, and was developed as an open source stack in \cite{2013bloesslOpenSourceIEEE}. In order increase the transmitted signal power of the SDR we added an signal amplifier and for time synchronization we are using a GPS disciplined oscillator that can be accessed by the SDR.\\

Following list gives an overview of the hardware elements of the experimental radio.
\begin{itemize}
	\item \textbf{Software Defined Radio:} Ettus USRP B210
	\item \textbf{Companion Computer:} Intel NUC (Ubuntu 20.04.6 LTS, GNU Radio)
	\item \textbf{Amplifier:} Coaxial ZX60-83LN 21~dB gain
	\item \textbf{Time Synchronization:} Board Mounted GPSDO (TCXO) 
	\item \textbf{Power Supply:} 100~W DC Converter for 19V, 100~W DC Converter for 6V
\end{itemize}
The elements are also illustrated in the payload setup shown in fig.~\ref{fig:payload}.

The GNU radio implementation uses a TAP interface, which is a virtual Ethernet device, on the companion computer and it enables to use the SDR as an IP based data link device. Thus, the experimental radio is able to transmit different application data via IP interface. As first modifications towards DroneCAST we also changed the center frequency to 5050~MHz, which is within the foreseen frequency band of 5030~MHz to 5091~MHz and halved the bandwidth from 10~MHz to 5~MHz and tested the modifications.

\begin{figure}[H]
	\centering
	\includegraphics[width=0.98\columnwidth]{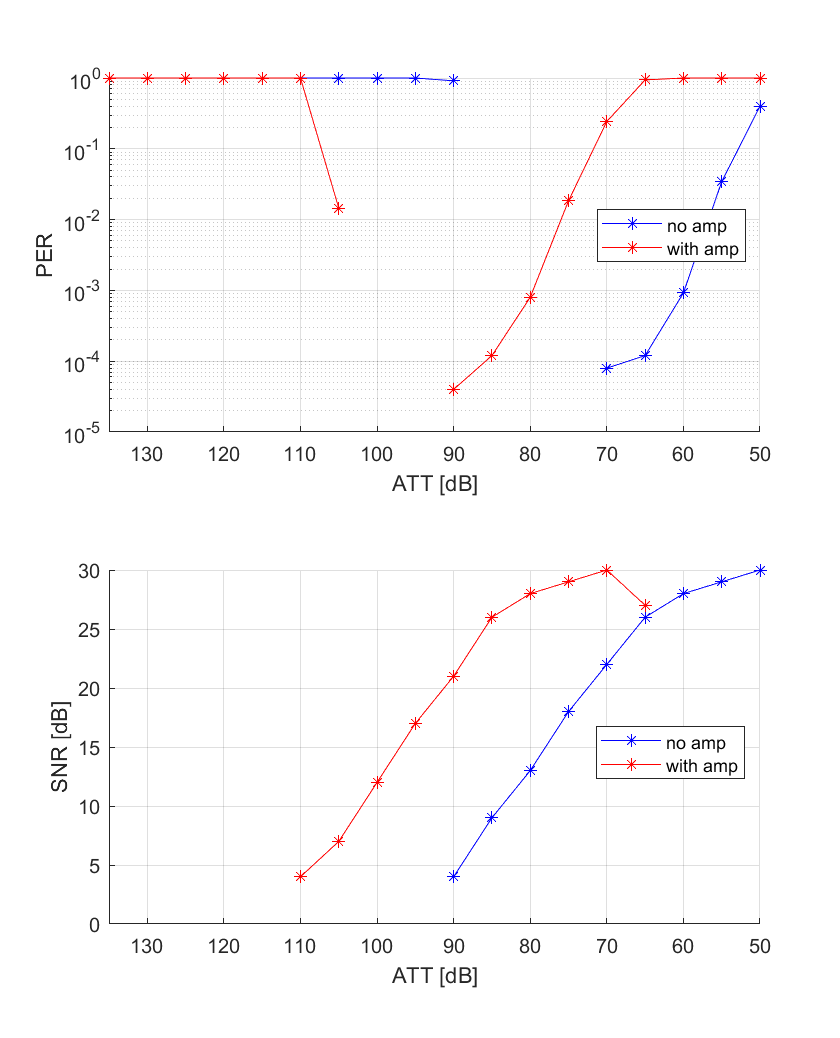}
	\caption{Measured packet error rates and received SNR values under ideal lab conditions without and without signal amplification.}
	\label{fig:labtest_sdr}
\end{figure}

We first evaluated the performance of our setup under ideal laboratory conditions by connecting two experimental radios with defined attenuators. Thereby we transmitted $25\cdot10^3$ packets of 125~Byte payload data with a transmission rate of 10~Hz for different attenuation values with and without amplifying the signal. Without amplification the USRP B210 is able to transmit with up to 10~dBm. With the amplifier we set the resulting transmission power to about 23~dBm in order to achieve similar transmission power as for the COTS radio, a Cohda Wireless Mk5. The amplifier has a gain of 21~dB, which means the same settings on the SDR result in a transmission power about 2~dBm. We repeated the measurement with and without the amplifier in order to evaluate if the amplifier causes signal distortions which would lead to increased packet errors. 
Figure~\ref{fig:labtest_sdr} shows the resulting packet error rates over different attenuation values. Furthermore, it shows the \ac{snr} values at packet reception indicated by the SDR. These values help us to evaluate the in-flight measurement results described in section~\ref{sec:in-flight_measurements}.
We can clearly see, that the two measurements with and without amplifier only differ according to the gain of 21~dB. Therefore, we assume a low-distortion amplification and the amplifier is feasible for our setup. Furthermore, the measurements reveal that the radios are only able to receive packets in a certain SNR range. For attenuation of 110~dB and 90~dB respectively the received signal power is too low and starting with 90~dB and 70~dB respectively the signal power starts to overdrive the radio until no packet can be decoded any more. The dynamic range is relatively small, because there is no active \ac{agc}. The hardware offers a built-in \ac{agc}, but the hardware driver does not active it when using the GNU radio framework.
%

\subsection{Integrated Payload on Hexacopter and Flighttrial Setup}
\begin{figure}[H]
	\centering
	\includegraphics[width=0.98\columnwidth]{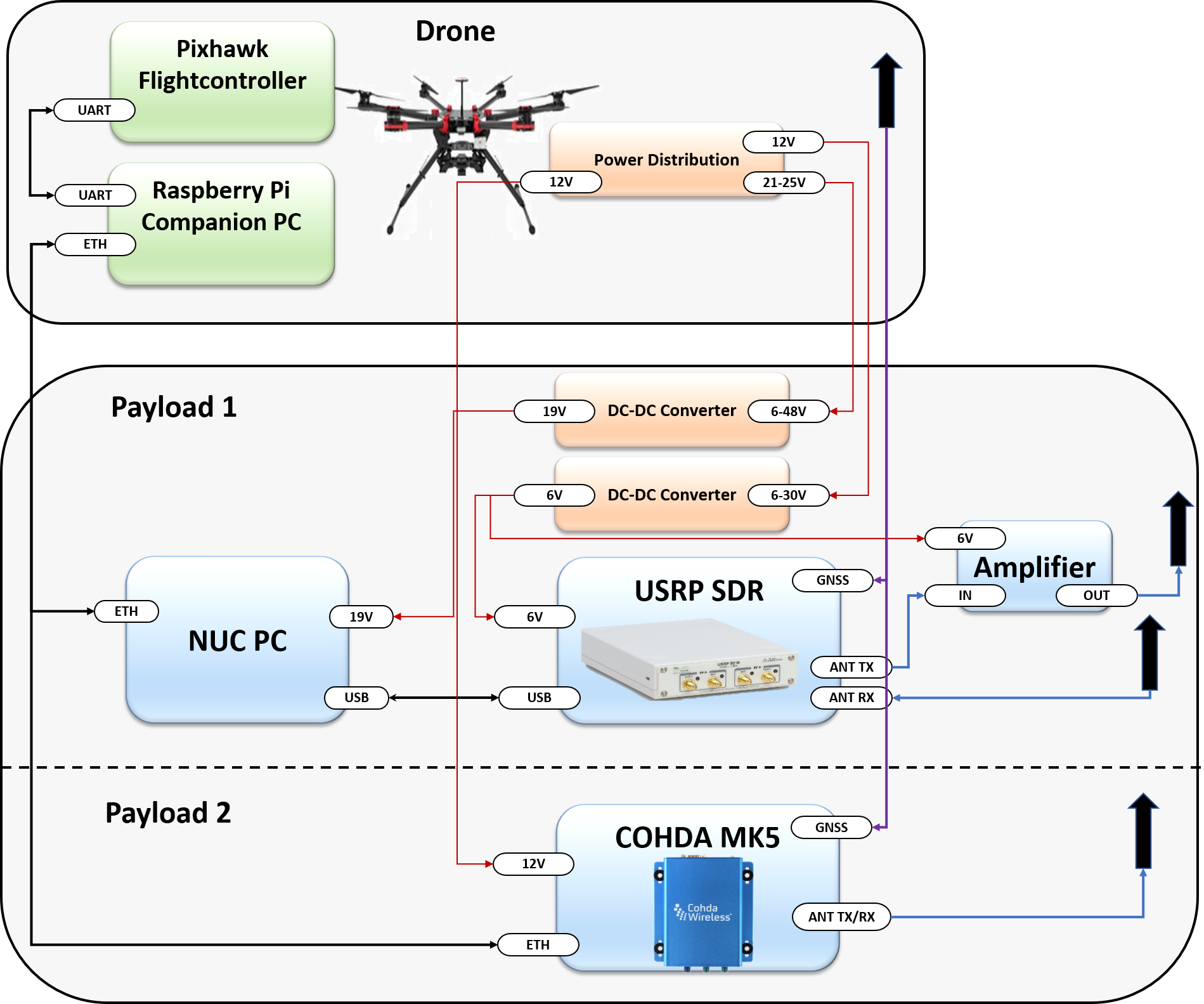}
	\caption{Overview of elements for integrated paylaod on drones.}
	\label{fig:payload}
\end{figure}

We integrated our experimental radio as payload on our hexacopters. We are using two custom build hexacopters based on DJI S900 airframes with upgraded E1200 propulsion system. They are equipped with two 10~Ah batteries at nominal voltage of 22.2~V and are able to carry up to 3~kg of payload with a flight time of approximately 15 to 20 minutes. Furthermore, they are using a Pixhawk 2 flight controller and have Raspberry Pi 4 companion computers in order to communicate with the flight controller via MAVLink messages over a serial interface. 
Figure~\ref{fig:payload} shows the main elements of our flight trial setup used for all in-flight measurements. Thereby we switched between the experimental radio and the COTS radio, but we were also able to carry both payloads for the flight demonstrations. 
This setup enables the transmission of flight controller data over the radios for different applications and also to send commands to the flight controller.

\subsection{In-flight Measurements} \label{sec:in-flight_measurements}
\begin{figure}[h!]
	\centering
	\includegraphics[width=0.98\columnwidth]{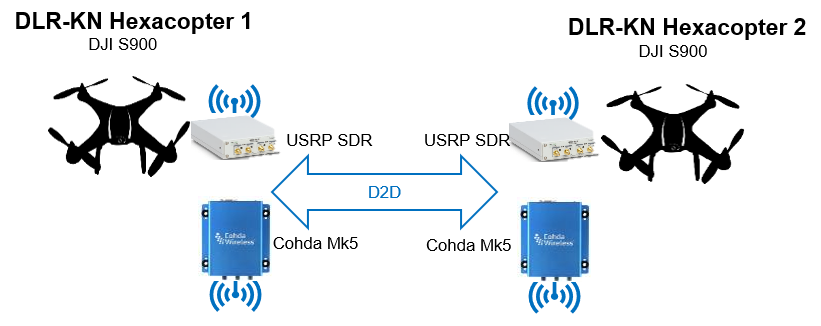}
	\caption{Overview of flight trial setup for the in-flight measurements.}
	\label{fig:setup_measurements}
\end{figure}

We equipped our two hexacopters with the given hardware prototypes for collecting in-flight measurement data in and around the model city. In order to asses the experimental radio, we first performed three different flight missions and compared the performance between the experimental radio and the COTS hardware. Then we measured the performance when flying close and in between the model city providing nonLOS scenarios. Figure~\ref{fig:setup_measurements} illustrates the overall flight trial setup for the measurements consisting of our two hexacopters equipped with one of the two payload options. 

The measurements and results are discussed in the following sections.
Overall the measurements and demonstrations showed a feasible bidirectional information exchange for the experimental radio setup. We increased the transmission power by using an signal amplifier in order to achieve similar transmission power compared to the COTS radio and we used additional GPSDO extension boards in order to synchronize all the radios with GPS time reference. However, the performance of our experimental radio was slightly worse to the COTS hardware due to a missing \ac{agc} and weaker SNR values. In all \ac{los} measurement scenarios, no packet losses occurred for the COTS radio whereas the experimental radio showed packet errors between $4\% - 8\%$.

\subsubsection{Flight Mission 1}
\begin{figure}[h!]
	\centering
	\subfloat[2D]{\includegraphics[width=0.98\columnwidth]{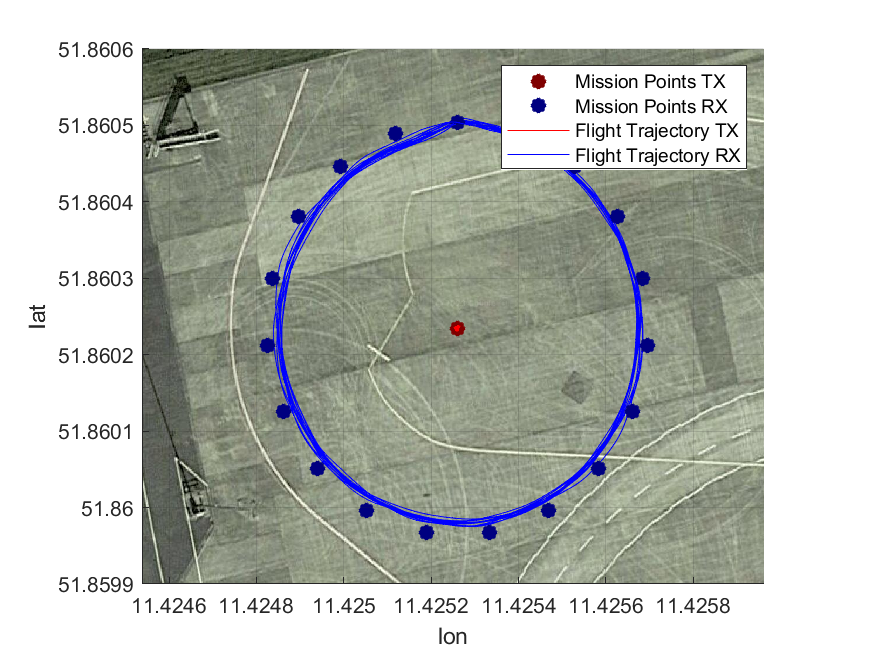}%
		\label{fig:mission1_2d}}
	\hfil
	\subfloat[3D]{\includegraphics[width=0.98\columnwidth]{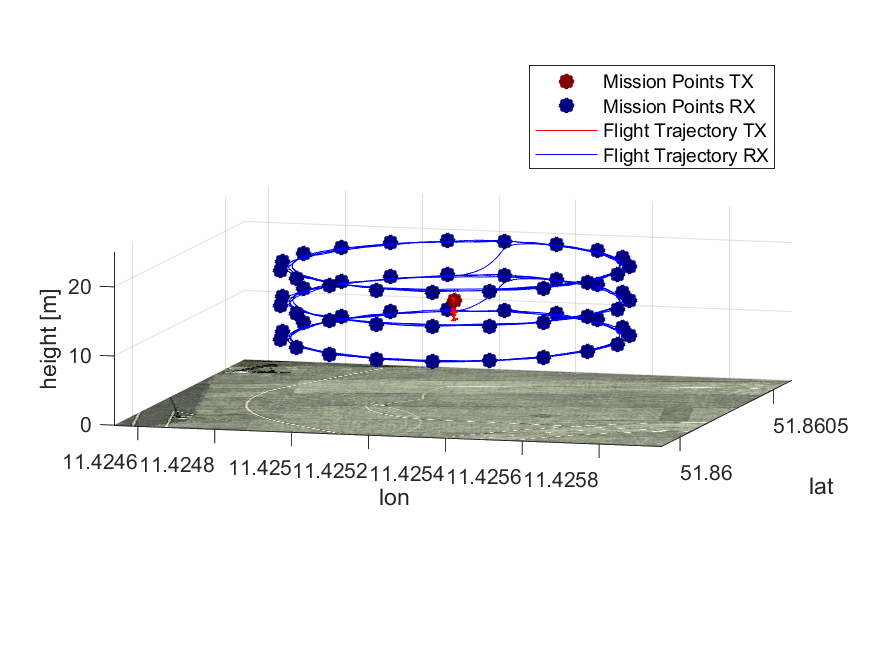}%
		\label{fig:mission1_3d}}
	\caption{Mission 1: Waypoint mission points and flight trajectories for transmitting (TX) and receiving (RX) drone.}
	\label{fig:mission1}
\end{figure}

\begin{figure}[h!]
	\centering
	\includegraphics[width=0.98\columnwidth]{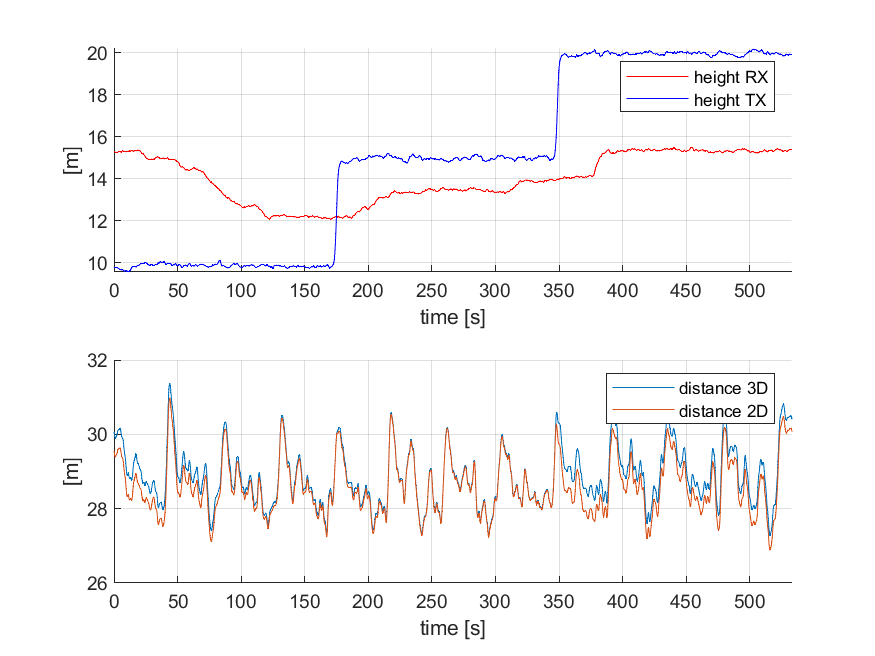}
	\caption{Mission 1: Heights of drones and distances between them.}
	\label{fig:mission1_height}
\end{figure}

For Mission~1 the transmitting drone was hovering at a defined height of 15~m and the receiving drone was flying four circles with a radius of 30~m around the hovering drone in three different heights at 10~m, 15~m and 20~m. Figure~\ref{fig:mission1} illustrates the flight mission and figure~\ref{fig:mission1_height} shows the flight heights as well as the distances between the drones as direct three-dimensional distance and as two-dimensional distance above the ground. The flying drone always headed towards the next mission waypoint indicated as dots in the figure. Due to navigation accuracy, the flown trajectories were not always the same but closely followed the path given the waypoints for all measurement scenarios. In this scenario, the distance between the drones stays more or less the same and only slightly changes with the drone heights. Therefore, this scenario enables to analyze the impact of air frame shadowing without changing the fading due to multipath propagation.

\subsubsection{Results Mission 1}
\begin{figure}[h!]
	\centering
	\includegraphics[width=0.98\columnwidth]{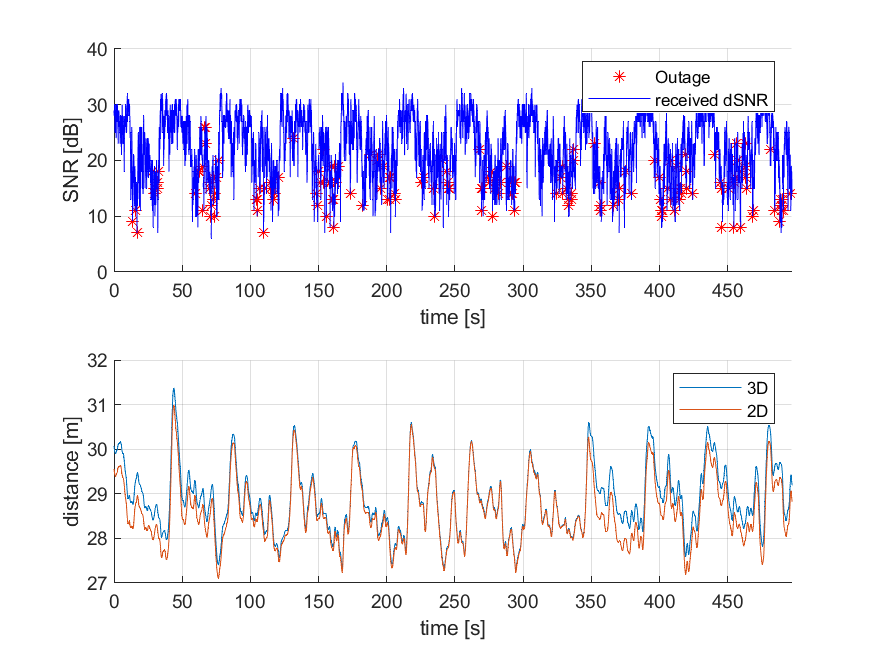}
	\caption{Mission 1: Measurement results for experimental radio setup.}
	\label{fig:mission1_sdr_res}
\end{figure}

\begin{figure}[h!]
	\centering
	\subfloat[2D]{\includegraphics[width=0.98\columnwidth]{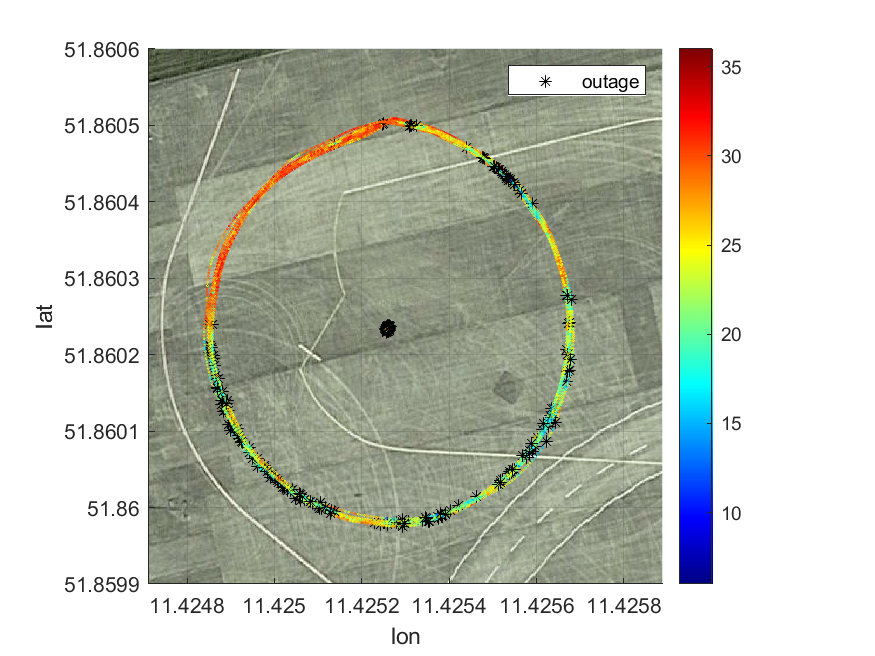}%
		\label{fig:mission1_sdr_2d}}\hfil
	\subfloat[3D]{\includegraphics[width=0.98\columnwidth]{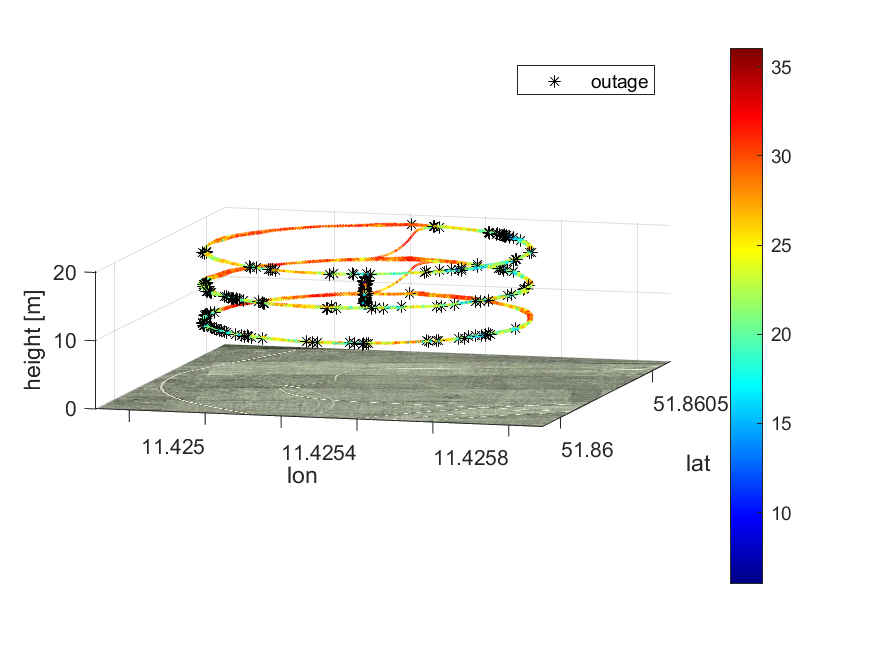}%
		\label{fig:mission1_sdr_3d}}
	\caption{Mission 1: Measurement results overlay-ed on map for experimental radio setup.}
	\label{fig:mission1_sdr_map}
\end{figure}

Figure~\ref{fig:mission1_sdr_res} shows the measured received SNR values of the experimental radio for the whole flight and the instants in time when packets where not successfully received together with the distances. It can be seen that mostly packet errors occur at low SNR values and the SNR values reveal a repeating pattern and vary between values lower than 10~dB and higher than 30~dB. 
Figure~\ref{fig:mission1_sdr_map} shows again the received SNR value and the packet errors overlay-ed on the trajectories in two and three dimensional layouts. On the map we can clearly see that the repeating pattern of the SNR values results from different height independent viewing angles between the drones and are caused by the airframe shadowing of the hovering drone.
The overall packet error rate for this measurement was about $4\%$ and the main reason was a too weak achieved SNR value for the experimental radio.

\begin{figure}[h!]
	\centering
	\includegraphics[width=0.98\columnwidth]{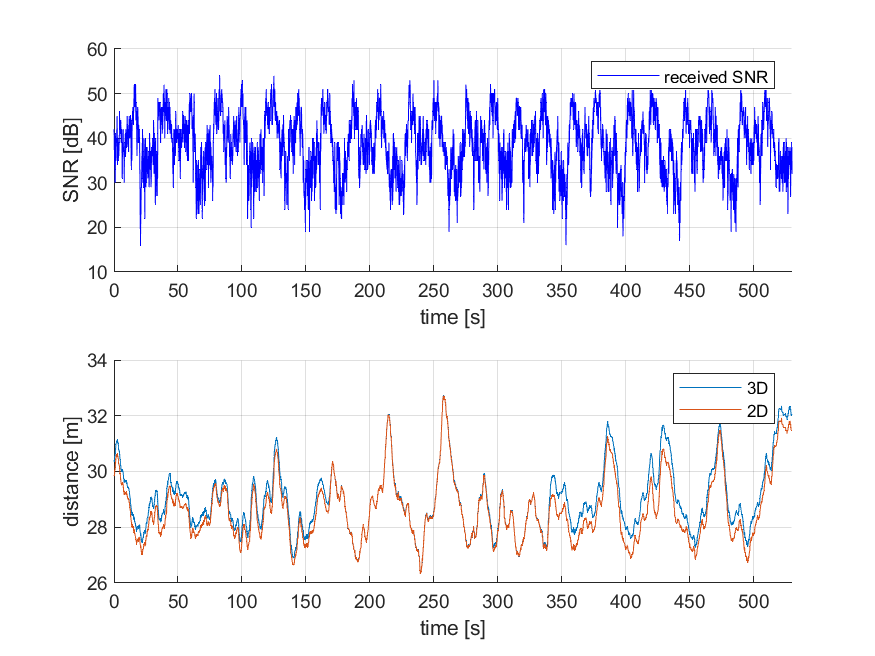}
	\caption{Mission 1: Measurement results for COTS hardware setup.}
	\label{fig:mission1_cohda_res}
\end{figure}

\begin{figure}[h!]
	\centering
	\subfloat[2D]{\includegraphics[width=0.98\columnwidth]{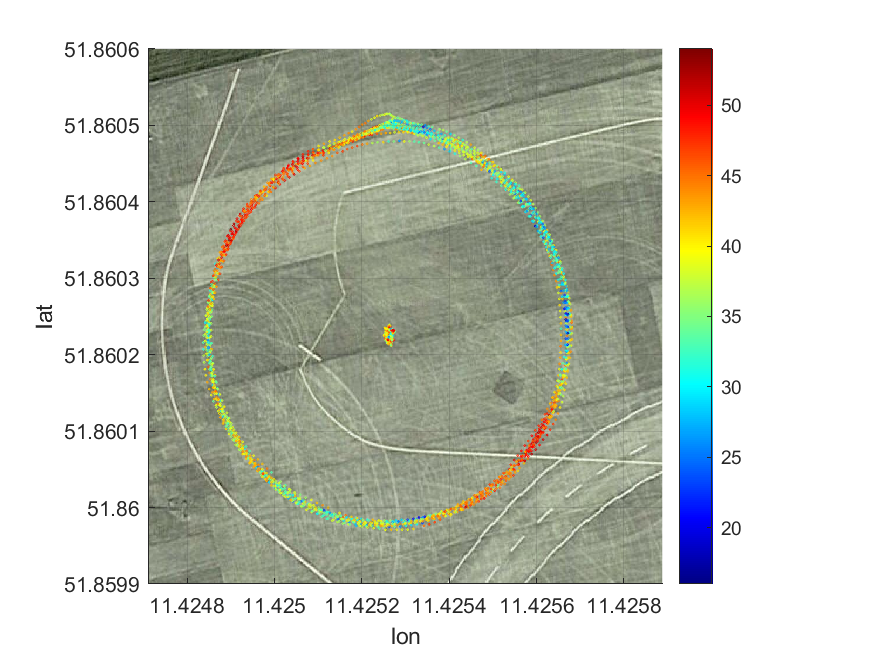}%
		\label{fig:mission1_cohda_2d}}
	\hfil
	\subfloat[3D]{\includegraphics[width=0.98\columnwidth]{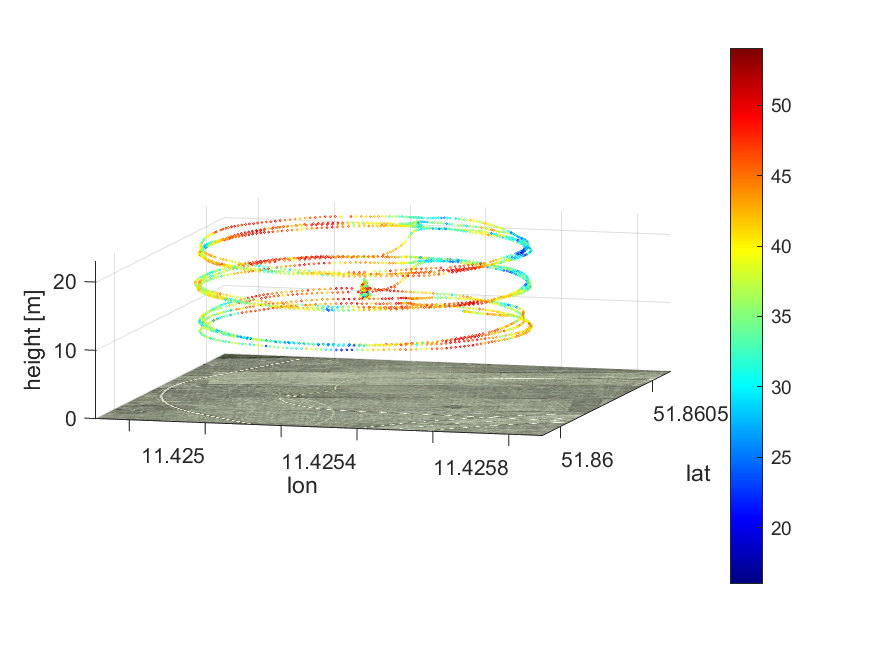}%
		\label{fig:mission1_cohda_3d}}
	\caption{Mission 1: Measurement results on map for COTS hardware setup.}
	\label{fig:mission1_cohda_map}
\end{figure}

For comparison to the experimental radio we performed this measurement for the COTS hardware. Figure~\ref{fig:mission1_cohda_res} shows the received SNR values and the distances, but this time no packet errors have occurred. The repeating pattern is again recognizable and the indicated values are higher compared to the values of the experimental radio but the differences are similar. This result shows, that the COTS radio can achieve higher SNR values for the same received signal powers.
Figure~\ref{fig:mission1_cohda_map} illustrates the results again on a map. We again can clearly see the influence of the airframe shadowing.

\subsubsection{Flight Mission 2}
\begin{figure}[h!]
	\centering
	\subfloat[2D]{\includegraphics[width=0.98\columnwidth]{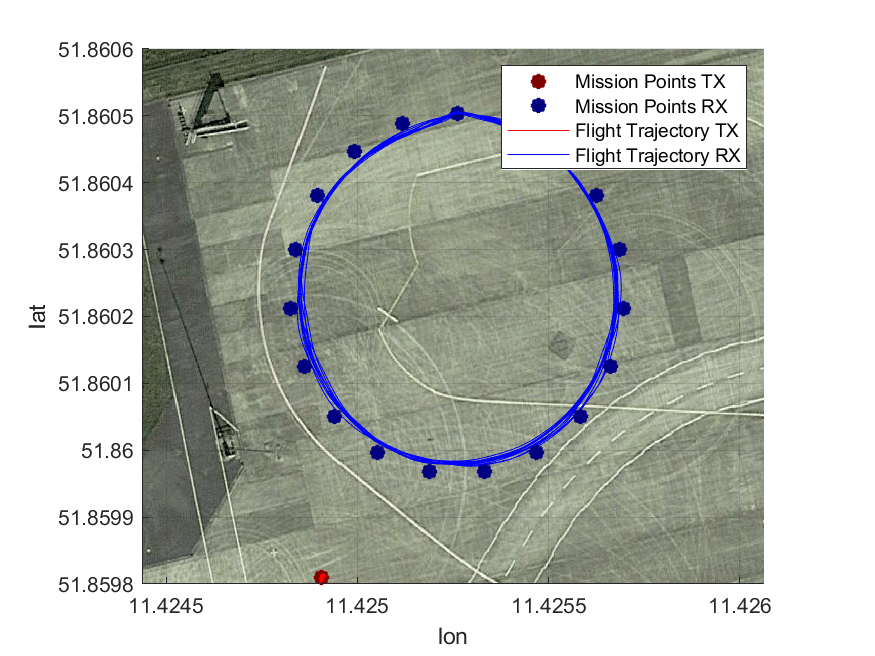}%
		\label{fig:mission2_2d}}
	\hfil
	\subfloat[3D]{\includegraphics[width=0.98\columnwidth]{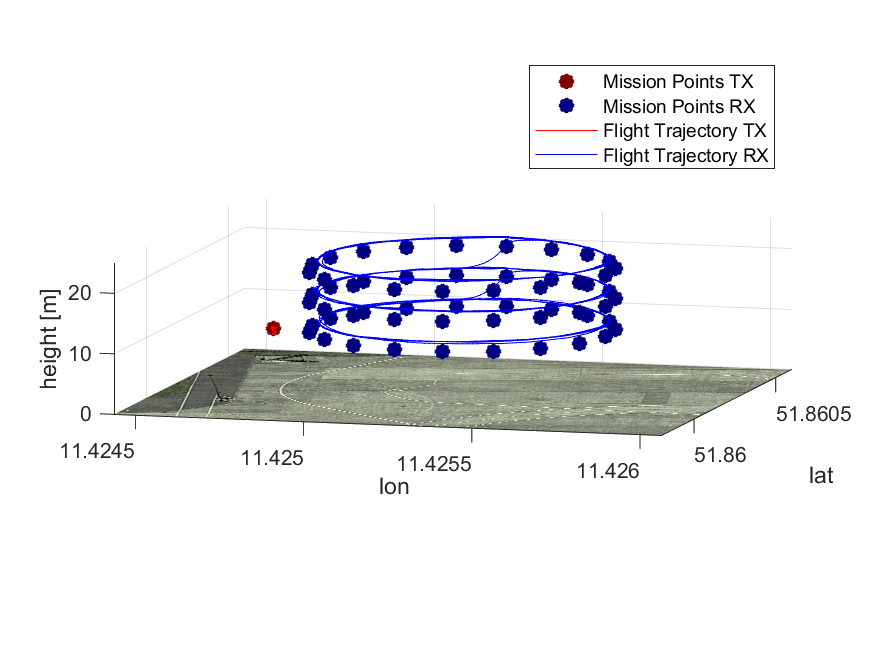}%
		\label{fig:mission2_3d}}
	\caption{Mission 2: Waypoint mission points and flight trajectories for transmitting and receiving drone.}
	\label{fig:mission2}
\end{figure}

\begin{figure}[h!]
	\centering
	\includegraphics[width=0.98\columnwidth]{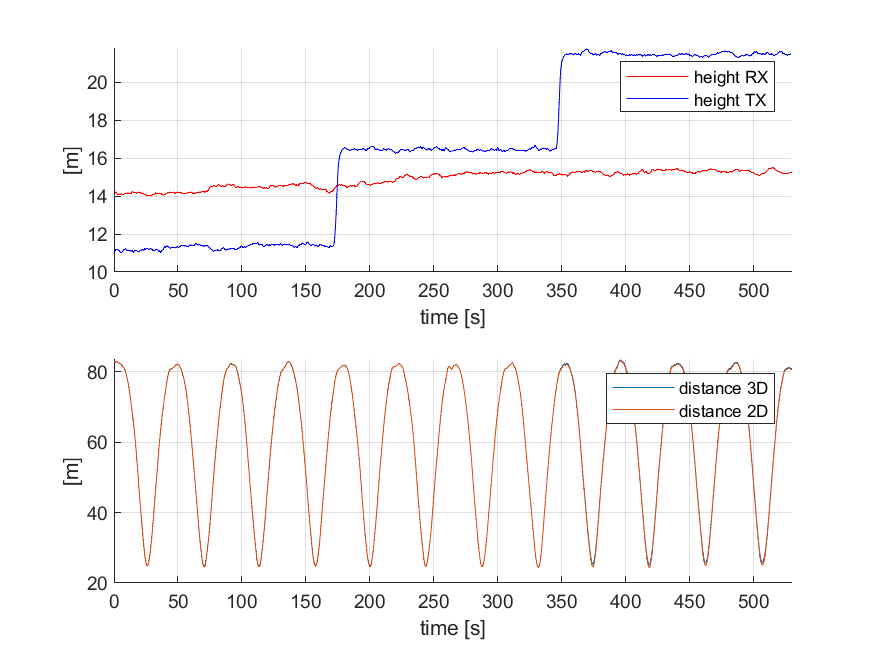}
	\caption{Mission 2: Heights of drones and distances between.}
	\label{fig:mission2_height}
\end{figure}

For Mission~2 the receiving drone flew the same flight trajectory as in Mission~1, but the transmitting drone was hovering at a different position outside the circles. In this scenario, the distance between the drones changes in order to analyze the influence of fading in comparison to the results of Mission~1. Figure~\ref{fig:mission2} illustrates the mission plotted on a map and fig.~\ref{fig:mission2_height} shows the drones heights and the distances between them.

\subsubsection{Results Mission 2}
\begin{figure}[h!]
	\centering
	\includegraphics[width=0.98\columnwidth]{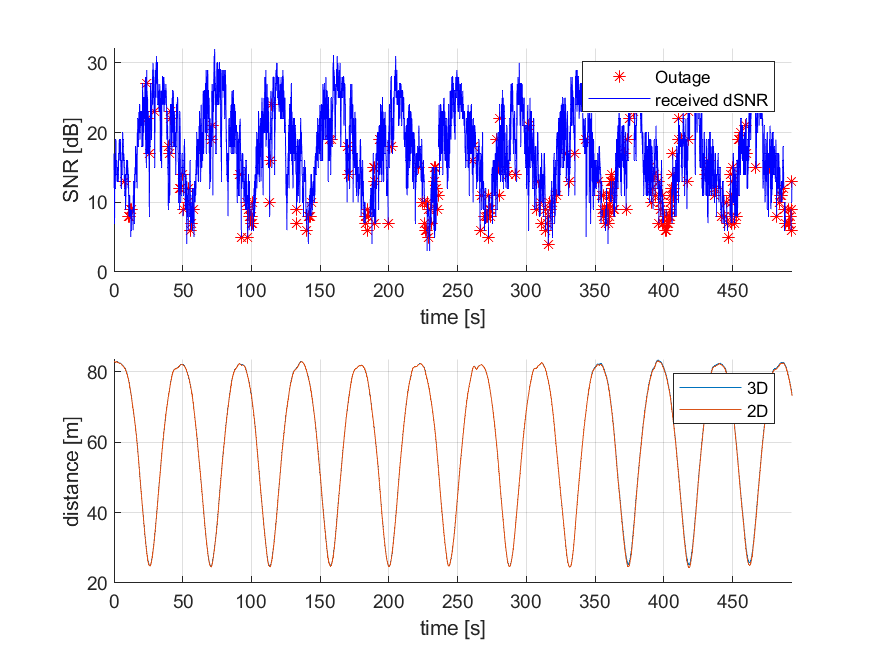}
	\caption{Mission 2: Measurement results for SDR setup.}
	\label{fig:mission2_sdr_res}
\end{figure}

\begin{figure}[h!]
	\centering
	\subfloat[2D]{\includegraphics[width=0.98\columnwidth]{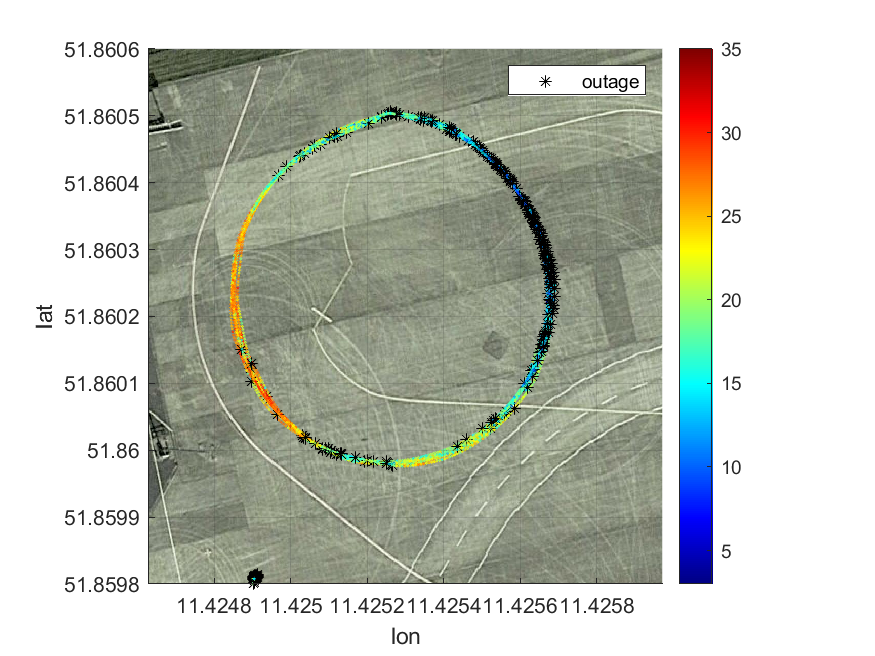}%
		\label{fig:mission2_sdr_2d}}
	\hfil
	\subfloat[3D]{\includegraphics[width=0.98\columnwidth]{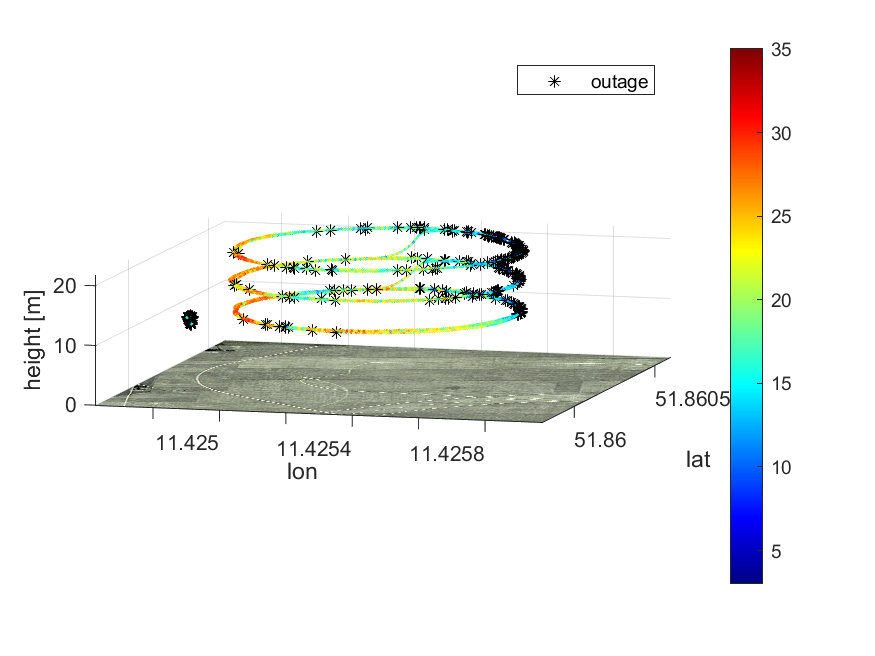}%
		\label{fig:mission2_sdr_3d}}
	\caption{Mission 2: Measurement results on map for SDR setup.}
	\label{fig:mission2_sdr_map}
\end{figure}

Figure~\ref{fig:mission2_sdr_res} shows the measured received SNR values of the experimental radio for the whole flight and the instants in time when packets where not successfully received together with the distances. It can be seen that mostly packet errors occur at low SNR values and the SNR values reveal a repeating pattern and vary between values lower than 10~dB and higher than 30~dB. 
Figure~\ref{fig:mission2_sdr_map} shows again the received SNR value and the packet errors overlay-ed on the trajectories in two and three dimensional layouts. On the map we can clearly see that the repeating pattern of the SNR values results from different height independent viewing angles between the drones and are caused by airframe shadowing but this time of the hovering drone and the flying drone. In comparison to Mission~1, slightly more packet errors occur.
The overall packet error rate for this measurement was about $8\%$ and the main reason was a too weak achieved SNR value for the experimental radio caused by airframe shadowing.

\begin{figure}[h!]
	\centering
	\includegraphics[width=0.98\columnwidth]{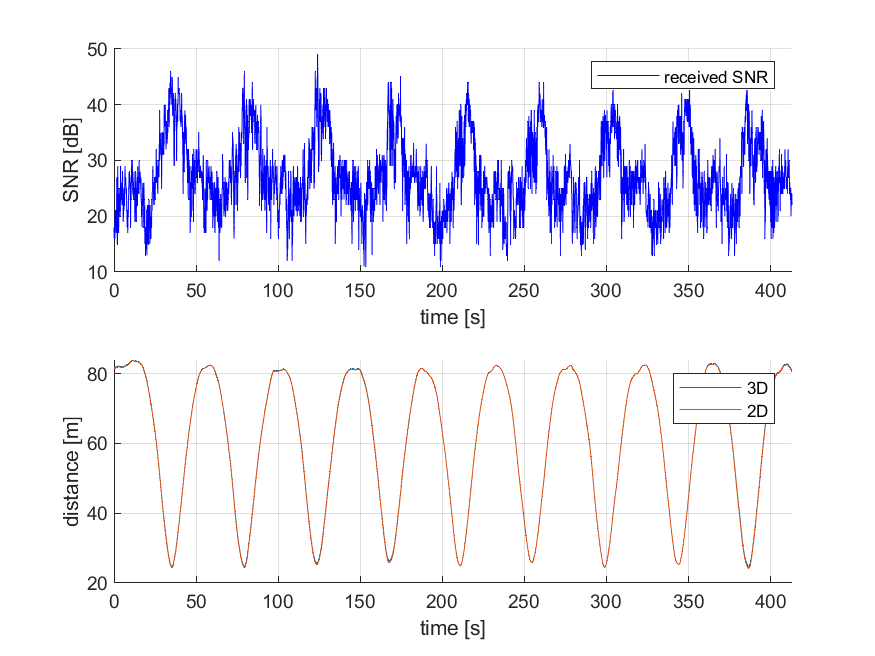}
	\caption{Mission 2: Measurement results for COTS hardware setup.}
	\label{fig:mission2_cohda_res}
\end{figure}

\begin{figure}[h!]
	\centering
	\subfloat[2D]{\includegraphics[width=0.98\columnwidth]{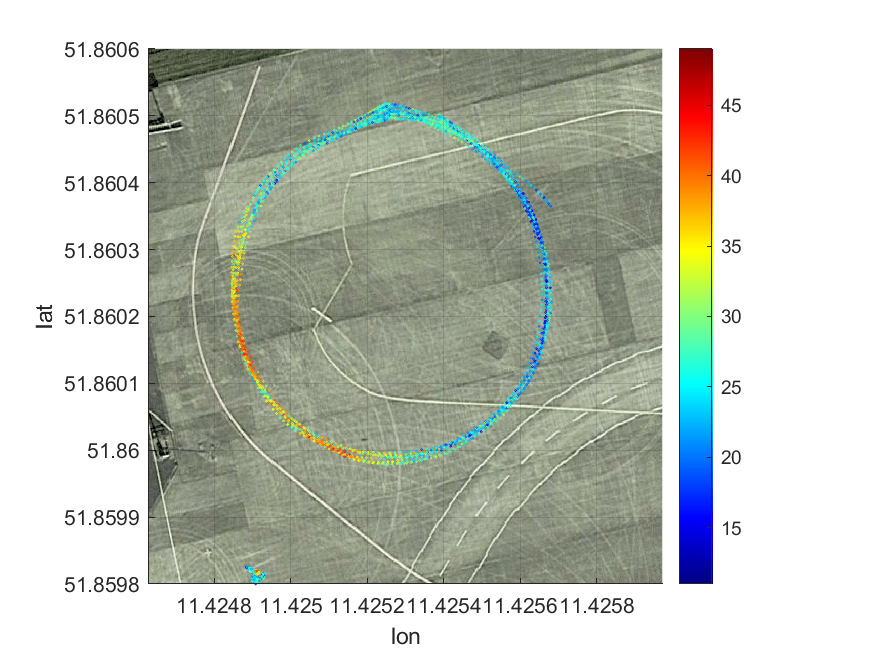}%
		\label{fig:mission2_cohda_2d}}
	\hfil
	\subfloat[3D]{\includegraphics[width=0.98\columnwidth]{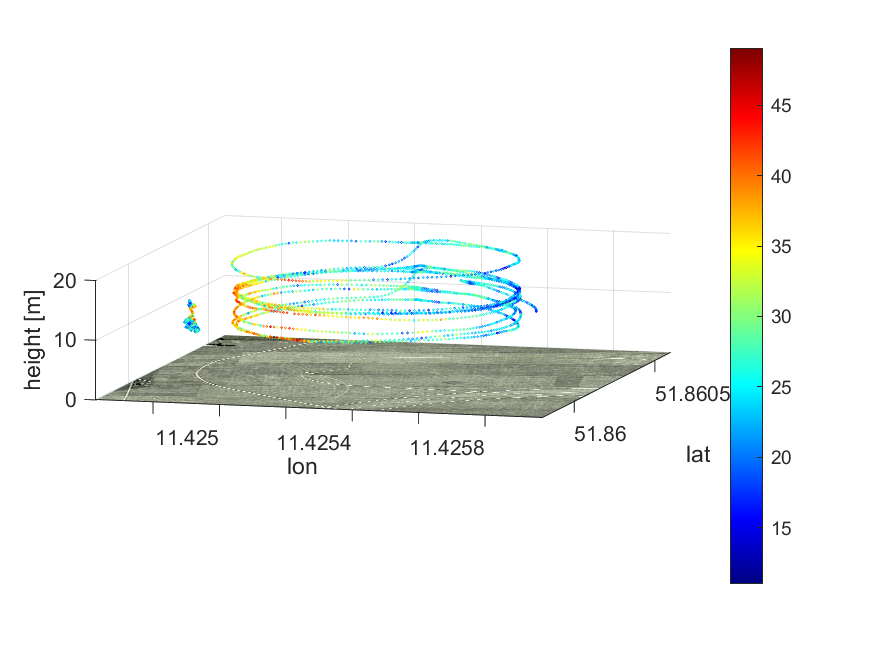}%
		\label{fig:mission2_cohda_3d}}
	\caption{Mission 2: Measurement results on map for COTS hardware setup.}
	\label{fig:mission2_cohda_map}
\end{figure}

For comparison to the experimental radio we performed this measurement for the COTS hardware. Figure~\ref{fig:mission2_cohda_res} shows the received SNR values and the distances, but no packet errors have occurred. The repeating pattern is again recognizable and the indicated values are higher compared to the values of the experimental radio but the differences are similar.
Figure~\ref{fig:mission2_cohda_map} illustrates the results again on a map. We again can clearly see the influence of the airframe shadowing.

\subsubsection{Flight Mission 3}
For Mission~3 both drones were flying at heights about 20~m above ground on parallel trajectories and were repeatingly coming close down to about 10~m distance and flying away from each other to about 60~m distance. For this scenario the viewing angles only changed when flying forwards or backwards the trajectory in order to analyze the influences of fading and distance without airframe shadowing.
\begin{figure}[h!]
	\centering
	\subfloat[2D]{\includegraphics[width=0.98\columnwidth]{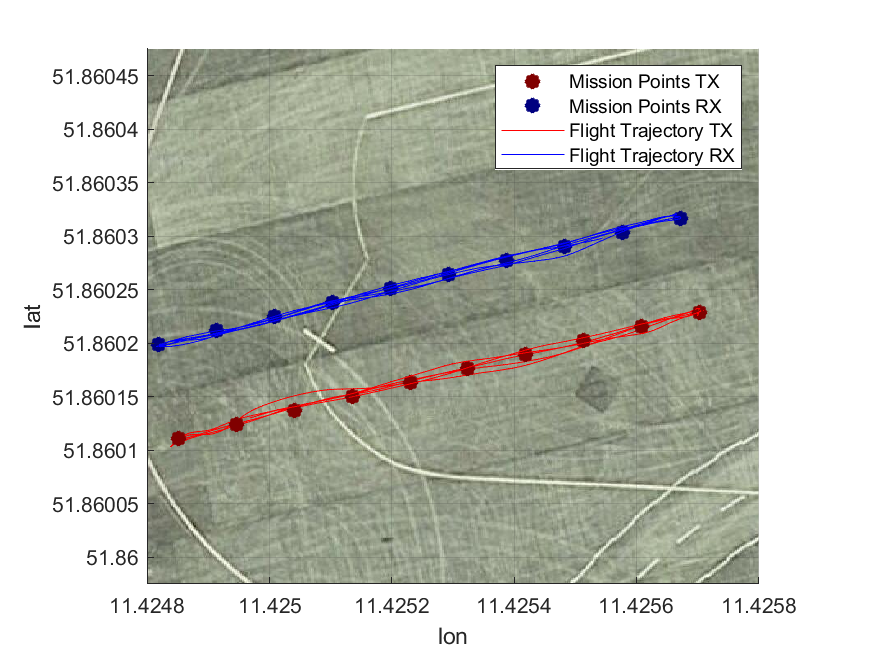}%
		\label{fig:mission3_2d}}
	\hfil
	\subfloat[3D]{\includegraphics[width=0.98\columnwidth]{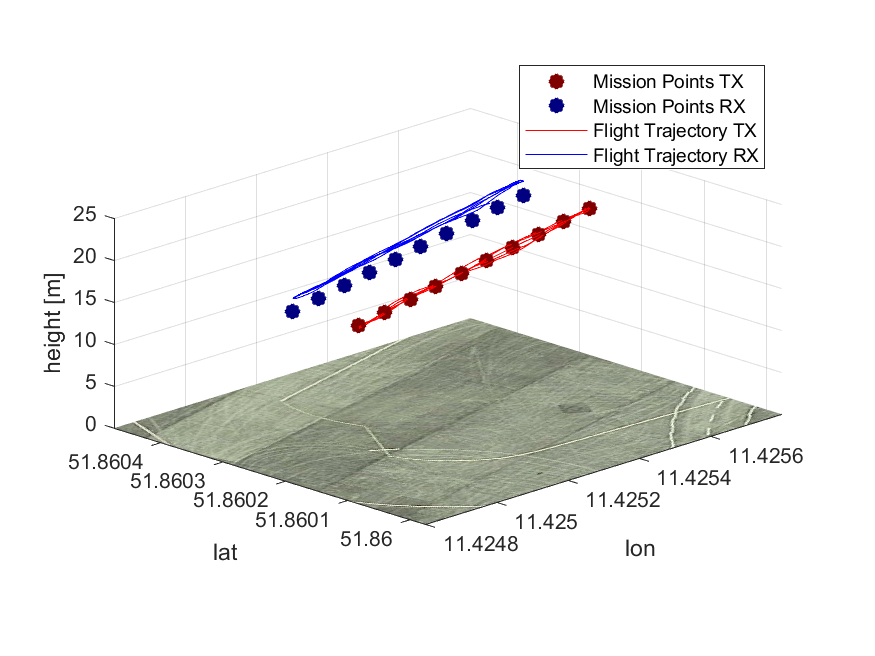}%
		\label{fig:mission3_3d}}
	\caption{Mission 3: Waypoint mission points and flight trajectories for transmitting and receiving drone.}
	\label{fig:mission3}
\end{figure}
\begin{figure}[h!]
	\centering
	\includegraphics[width=0.98\columnwidth]{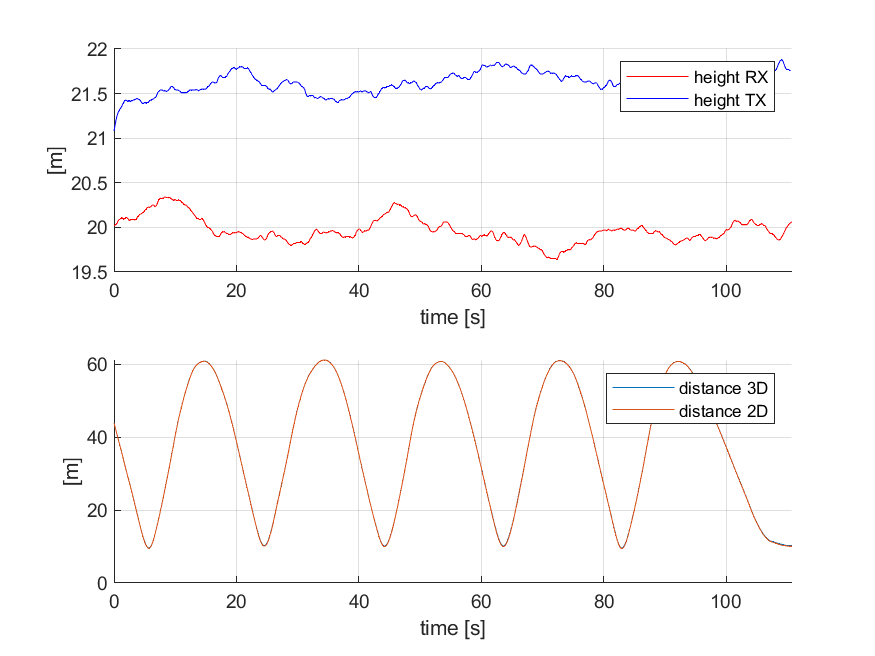}
	\caption{Mission 3: Heights of drones and distances between.}
	\label{fig:mission3_height}
\end{figure}

\subsubsection{Results Mission 3}
\begin{figure}[h!]
	\centering
	\includegraphics[width=0.98\columnwidth]{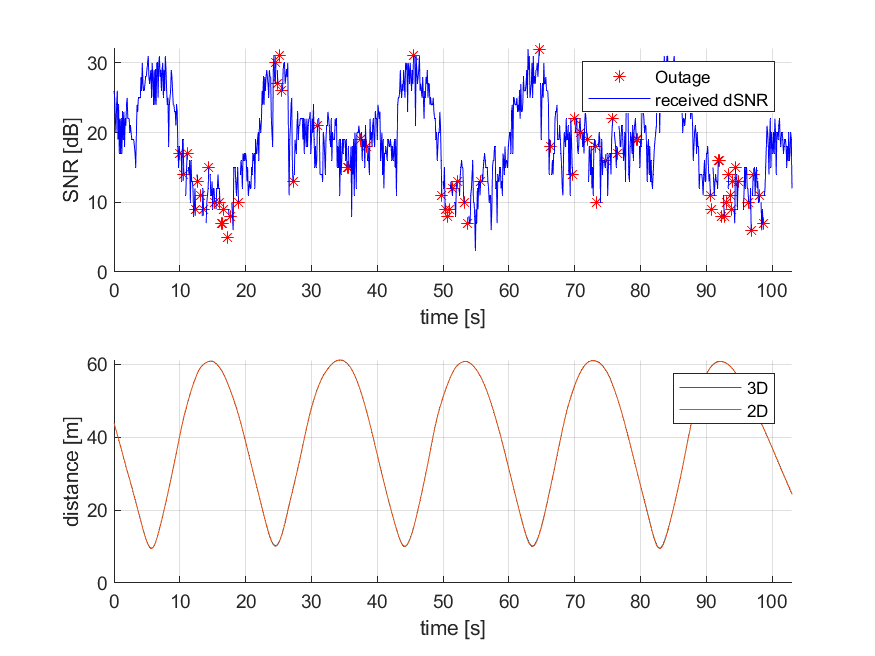}
	\caption{Mission 3: Measurement results for SDR setup.}
	\label{fig:mission3_sdr_res}
\end{figure}

\begin{figure}[h!]
	\centering
	\subfloat[2D]{\includegraphics[width=0.98\columnwidth]{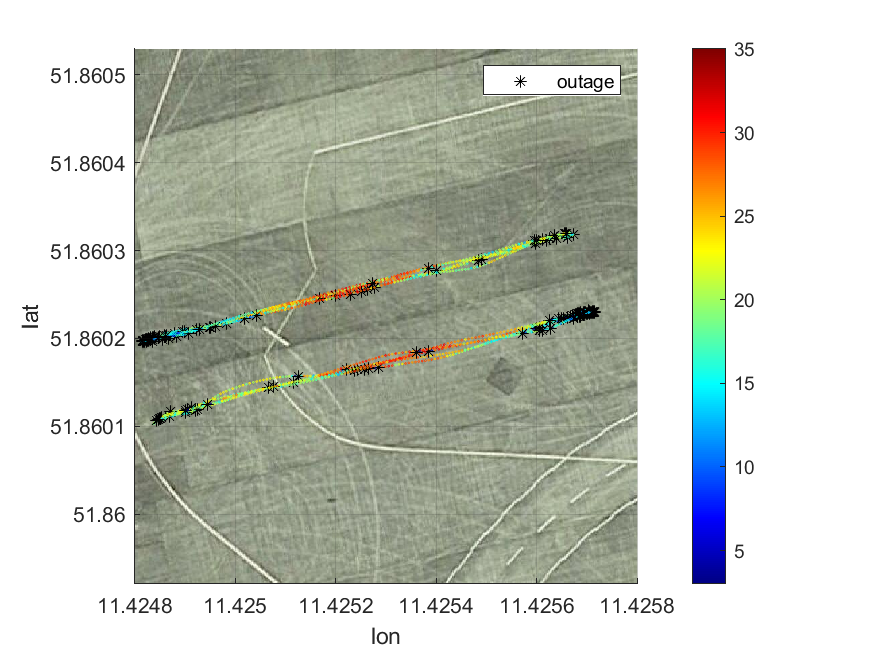}%
		\label{fig:mission3_sdr_2d}}
	\hfil
	\subfloat[3D]{\includegraphics[width=0.98\columnwidth]{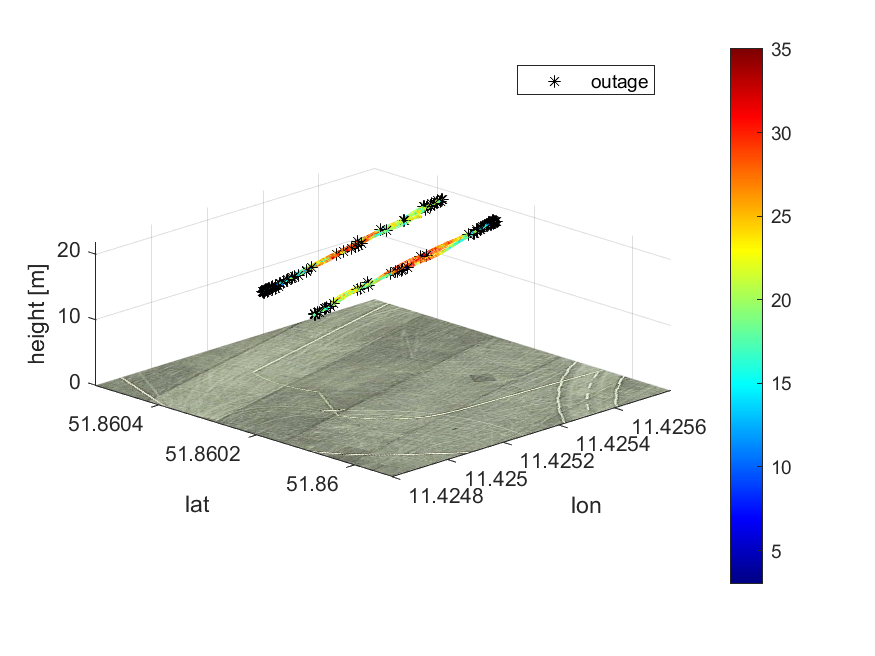}%
		\label{fig:mission3_sdr_3d}}
	\caption{Mission 3: Measurement results on map for SDR setup.}
	\label{fig:mission3_sdr_map}
\end{figure}

Figure~\ref{fig:mission3_sdr_res} shows the measured received SNR values of the experimental radio for the whole flight and the instants in time when packets where not successfully received together with the distances. This time we can see packet errors additionally due to high SNR values when the receiver was not able to handle the high signal power.   
Figure~\ref{fig:mission3_sdr_map} shows again the received SNR values and the packet errors overlay-ed on the trajectories in two and three dimensional layouts. On the map we can clearly see that the repeating pattern of the SNR values mostly results from the distances between the drones. When they come close to each other at half of the mission path then the received signal power gets too high due to low distance and when they are at furthest distance away from each other, the received signal power gets too low.
The overall packet error rate for this measurement was about $6\%$.

\begin{figure}[h!]
	\centering
	\includegraphics[width=0.98\columnwidth]{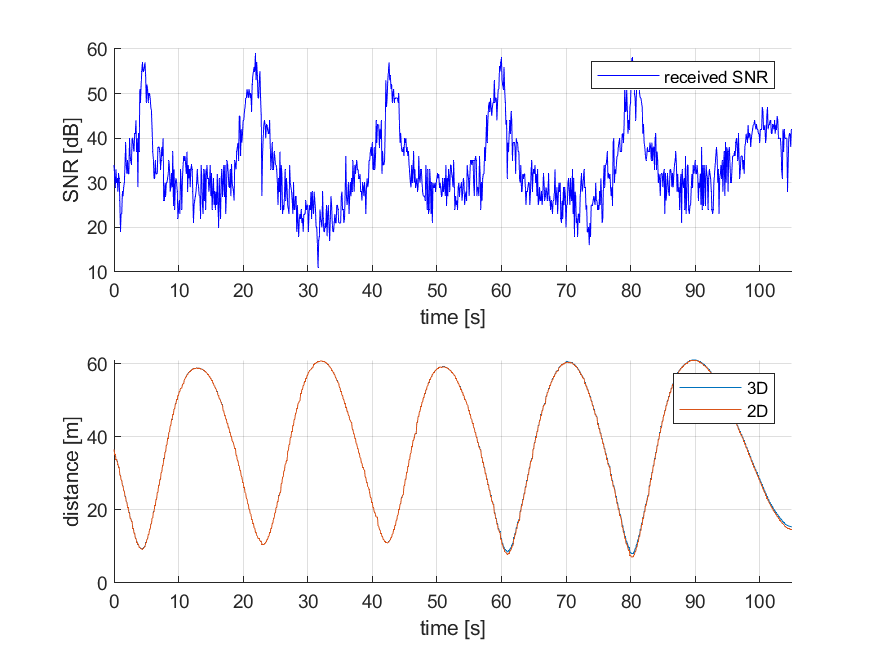}
	\caption{Mission 3: Measurement results for COTS hardware setup.}
	\label{fig:mission3_cohda_res}
\end{figure}

\begin{figure}[h!]
	\centering
	\subfloat[2D]{\includegraphics[width=0.98\columnwidth]{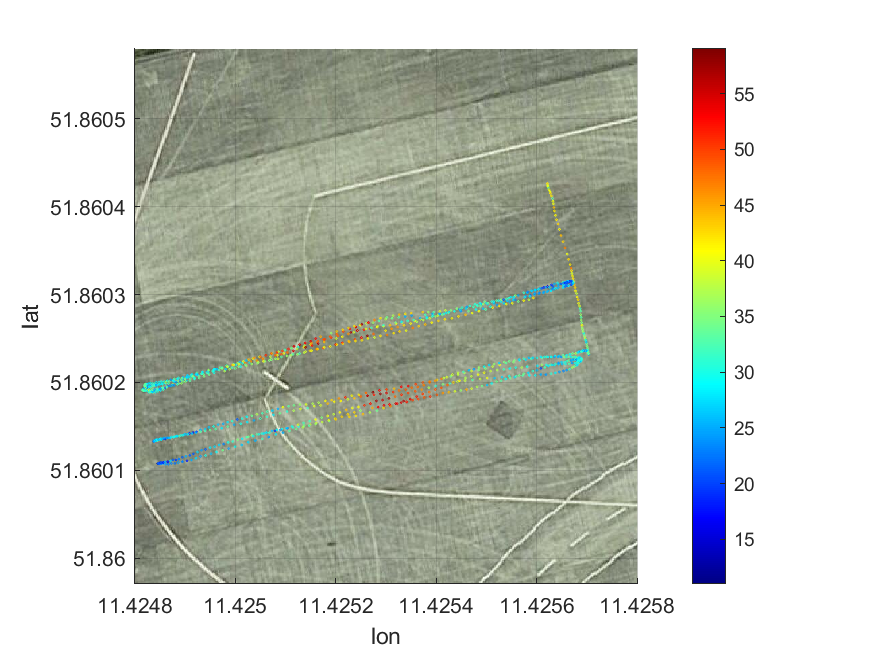}%
		\label{fig:mission3_cohda_2d}}
	\hfil
	\subfloat[3D]{\includegraphics[width=0.98\columnwidth]{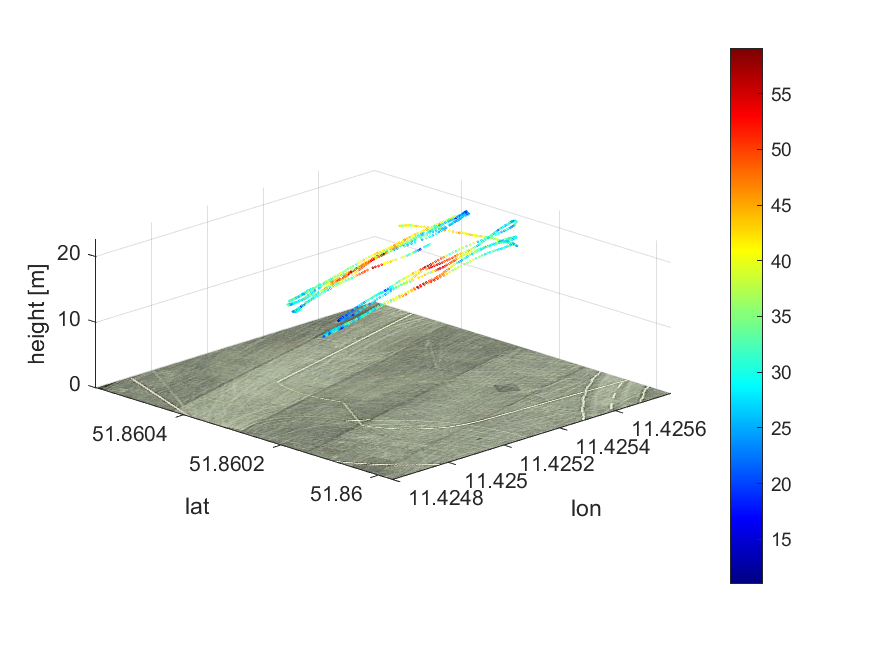}%
		\label{fig:mission3_cohda_3d}}
	\caption{Mission 3: Measurement results on map for COTS hardware setup.}
	\label{fig:mission3_cohda_map}
\end{figure}

For comparison to the experimental radio we performed this measurement for the COTS hardware. Figure~\ref{fig:mission3_cohda_res} shows the received SNR values and the distances, but no packet errors have occurred. The repeating pattern is again recognizable and the indicated values are higher compared to the values of the experimental radio but the differences are similar.
Figure~\ref{fig:mission3_cohda_map} illustrates the results again on a map. We again can clearly see the influence of the airframe shadowing.

\subsection{Flight Demonstrations at Model City}
\begin{figure}[H]
	\centering
	\includegraphics[width=0.99\columnwidth]{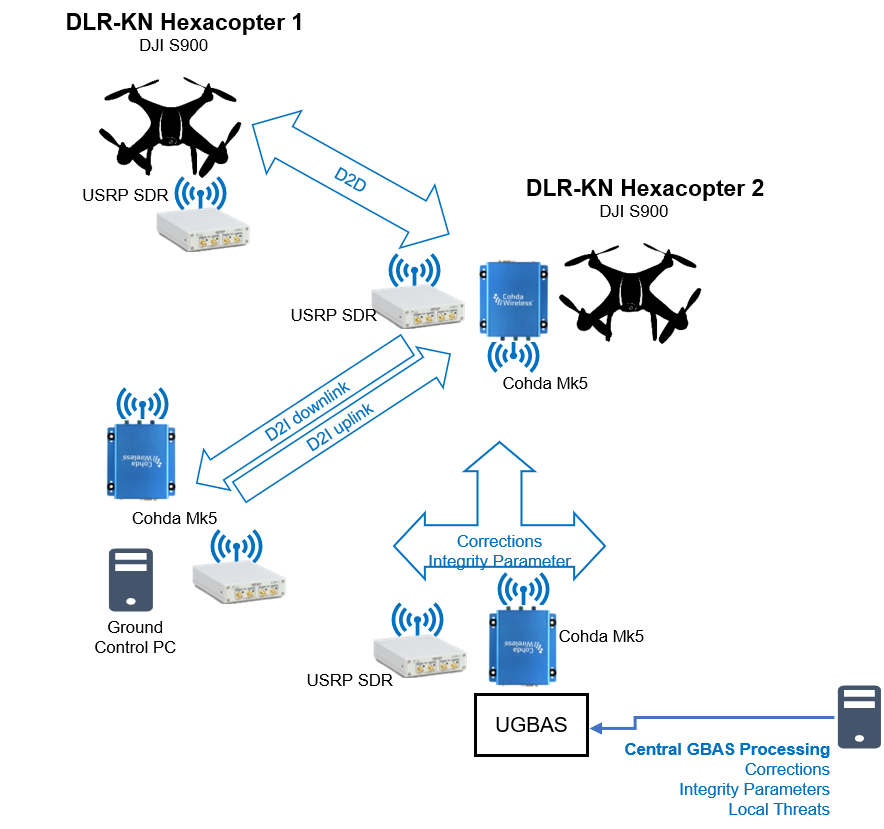}
	\caption{Overview of major elements of flight trial setup.}
	\label{fig:flighttrial_setup}
\end{figure}

In addition to our measurements, we also demonstrated following different drone-to-drone and drone-to-infrastructure communications around the model city
\begin{itemize}
	\item Secured Transmission of differential ground augmentation data (GBAS) with broadcast authentication protocol TESLA~\cite{2003perrigTESLABroadcastAuthentication}
	\item Transmission of the drone’s live position and monitoring on a ground station
	\item Cooperative collision avoidance by broadcasting flight trajectories and automatically stopping the drones to hold position
\end{itemize}
For this we used the overall setup shown in fig.~\ref{fig:flighttrial_setup}. For the collision avoidance scenario we used our experimental radio for the \ac{d2d} communication and transmitted monitoring messages down to a ground station via the COTS radio. In order to exchange trajectory information and to send commands to the flight controllers in order to stop their flight missions in case of emergent collision, we executed a python script on the raspberry pi companion computers on the drones and used the pymavlink library in order to communicate with the flight controller via MAVLink.

\begin{figure}[H]
	\centering
	\includegraphics[width=0.98\columnwidth]{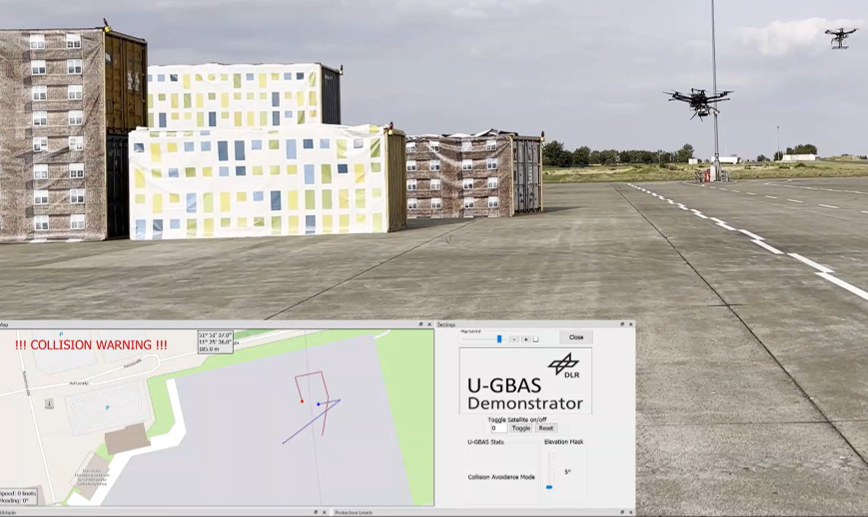}
	\caption{Flight demonstration of collision avoidance with experimental setup at model city.}
	\label{fig:demo_ca}
\end{figure}

\begin{figure}[H]
	\centering
	\includegraphics[width=0.98\columnwidth]{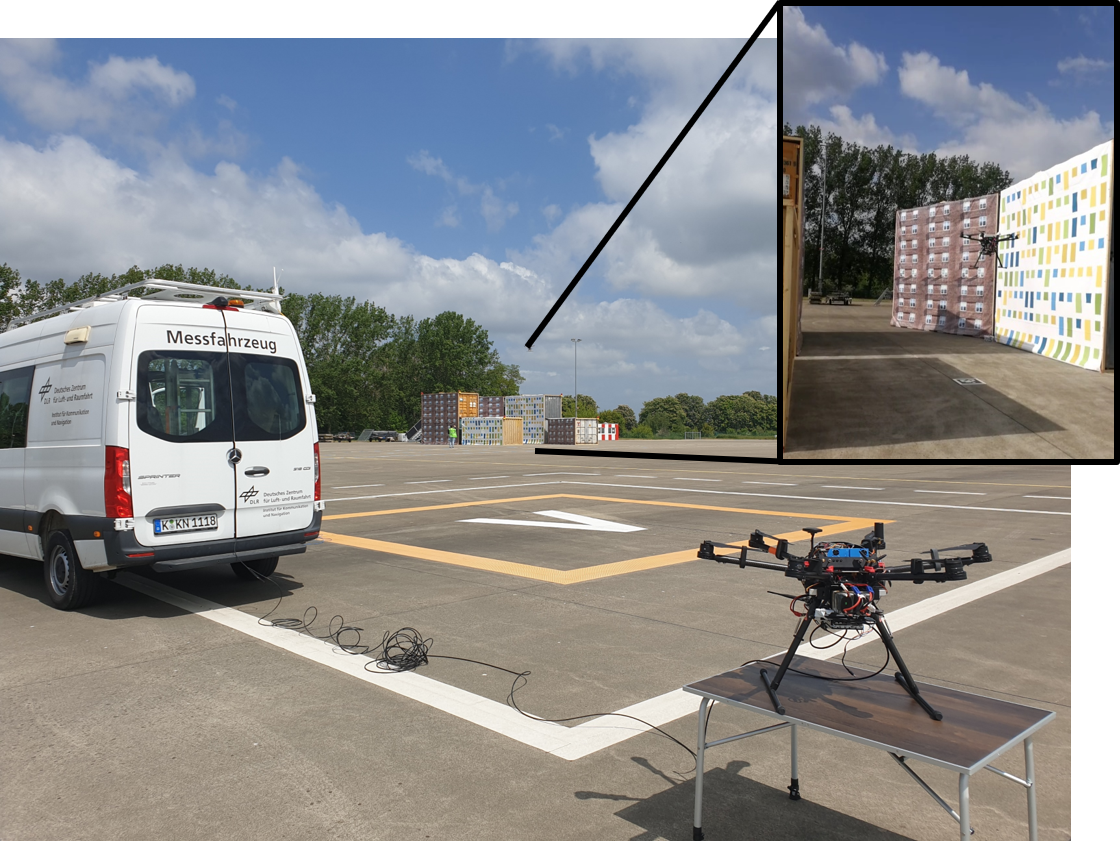}
	\caption{Flight demonstration of secured GBAS transmission to drone around model city.}
	\label{fig:demo_gbas}
\end{figure}

Figure.~\ref{fig:demo_ca} shows a video screenshot of the collision avoidance scenario demonstrated with our experimental setup at the model city. Thereby it shows the Live Monitoring screen with the received trajectories of both drones at a time when the collision avoidance application stopped the drones flight and sent an emergency message down to the monitoring ground station.
For the GBAS transmission for the drones we used one drone as broadcasting station at the ground like shown in fig.~\ref{fig:demo_gbas} and let one drone fly in and around the model city. In order to secure the transmission we used an software implementation of the TESLA protocol that was already demonstrated in our group at a flight trial with a piloted aircraft to secure the broadcast of GBAS correction data via LDACS \cite{2021maurerFlightTrialDemonstrationa}.




\section{Conclusion and Outlook}
In this work, we presented a multi-link approach with a focus on an ad-hoc communication concept that will help to reduce the probability of mid-air collisions and thus increase social acceptance of urban air mobility. As an essential part of the described multi-link approach, we aim to develop DroneCAST, a data link tailored to the special requirements and challenges in urban airspace, in order to establish an additional, decentralized and robust safety layer for the UTM concept. For the development of DroneCAST, we make use of our Drone-to-Drone Channel Model for Urban Environments, which is based on measurements in order to increase the robustness and efficiency.
As a first step towards an implementation, we equipped two drones with hardware prototypes of an experimental communication system and performed several flights around a model city to evaluate the performance of the hardware and to demonstrate different applications that will rely on robust and efficient communications. Results showed the feasibility of the experimental hardware setup. However, a missing automatic gain control for this setup resulted in a weaker performance compared to a COTS radio. Therefore, in the next steps we aim to develop a next level hardware prototype for a DroneCAST radio and also will target physical layer robustness and security topics.





\ifCLASSOPTIONcaptionsoff
  \newpage
\fi


\bibliographystyle{IEEEtran}
\bibliography{CEAS_Journal}



\end{document}